\newcommand{\nn}{\nonumber}

\newcommand{\cH}{{\cal H}} % Hilbert space
\newcommand{\cC}{{\cal C}} % conjugacy class
\newcommand{\cN}{{\cal N}} % centralizer group
\newcommand{\cF}{{\cal F}} % Free energy
\newcommand{\cV}{{\cal V}} % Heck operator
\newcommand{\cX}{{\cal X}} %
\newcommand{\cA}{{\cal A}} % long string oscillator
\newcommand{\ctA}{\widetilde{\cal A}} % long string oscillator
\newcommand{\cZ}{{\cal Z}} % Partition function for long string

\newcommand{\bR}{{\bf R}} % Real number
\newcommand{\bX}{{\bf X}} % String embedding
\newcommand{\tbX}{\widetilde{\bf X}} %Fourier transform
\newcommand{\bbX}{\overline{\bf X}} %Closed string variable
\newcommand{\bY}{{\bf Y}} % String embedding
\newcommand{\bZ}{{\bf Z}} % Integer
\newcommand{\bx}{{\bf x}} % transverse coordinate
\newcommand{\Bket}[2]{{|\,B:#1,#2\rangle\!\rangle}}
\newcommand{\Bbra}[2]{{\langle\!\langle B:#1,#2\,|}}
\newcommand{\Cket}[1]{{|\,C:#1^2,#1\rangle\!\rangle}}
\newcommand{\Cbra}[1]{{\langle\!\langle C:#1^2,#1\,|}}
\newcommand{\Ckets}[2]{{|\,C:#1,#2\rangle\!\rangle}}

\newcommand{\tsigma}{{\overline\sigma}} %for closed string world sheet
\newcommand{\tn}{{\overline{n}}}
\newcommand{\tm}{{\overline{m}}}
\newcommand{\tp}{{\overline{p}}}
\newcommand{\tk}{{\overline{k}}}

\newcommand{\cycl}{{\cal T}}
\newcommand{\cyclb}{{\overline{\cal T}}}
\newcommand{\inv}{{\cal S}}

\newcommand{\be}[1]{{\bf e}\!\left[#1\right]}
\newcommand{\qed}{{\em QED}}

\documentclass[12pt]{article}
\usepackage{epsf}
\usepackage{amssymb}

\catcode`\@=11
\@addtoreset{equation}{section}

\catcode`@=12

\begin{document}
%%%%%%%%%%%%%%%%%%%%%%%%%%%%%%%%%%%%%%%%%%%%%%%%%%%%%%%%%%%
%                                                         %
%  Title page                                             %
%                                                         %
%%%%%%%%%%%%%%%%%%%%%%%%%%%%%%%%%%%%%%%%%%%%%%%%%%%%%%%%%%%
\renewcommand{\thefootnote}{\fnsymbol{footnote}}
\font\csc=cmcsc10 scaled\magstep1
{\baselineskip=14pt
 \rightline{
 \vbox{\hbox{UT-888}
       \hbox{May 2000}
}}}

\vfill
\begin{center}
{\Large\bf
Open String on Symmetric Product
}
% \\ 
% \vskip 1mm
% {\large\em --- Toward second quantization of open string with D-branes ---}

\vfill

{\csc Hiroyuki Fuji}\footnote{
      e-mail address : fuji@hep-th.phys.s.u-tokyo.ac.jp} and 
{\csc Yutaka Matsuo}\footnote{
      e-mail address : matsuo@hep-th.phys.s.u-tokyo.ac.jp}\\
\vskip.1in

{\baselineskip=15pt
\vskip.1in
  Department of Physics,  Faculty of Science\\
  Tokyo University\\
  Bunkyo-ku, Hongo 7-3-1, Tokyo 113-0033, Japan
\vskip.1in
}

\end{center}
\vfill

\begin{abstract}
{%\baselineskip 15pt
We develop some basic properties of the open string on 
the symmetric product which is supposed to describe the 
open string field theory in discrete lightcone quantization
(DLCQ). After preparing the consistency conditions of the
twisted boundary conditions for Annulus/M\"obius/Klein Bottle
amplitudes in generic non-abelian orbifold, we classify the most 
general solutions of the constraints when the discrete group is $S_N$.
We calculate the corresponding orbifold amplitudes 
from two viewpoints -- from the boundary state formalism
and from the trace over the open string Hilbert space.
It is shown that the topology of the world sheet for the short
string and that of the long string in general do not coincide.
For example the annulus sector for the short string
contains all the sectors (torus, annulus, Klein bottle,
M\"obius strip) of the long strings. The boundary/cross-cap states 
of the short strings are classified into three categories in terms 
of the long string, the ordinary boundary and the cross-cap states, 
and the ``joint'' state which describes the connection of two short 
strings. We show that the sum of the all possible boundary conditions 
is equal to the exponential of the sum of the irreducible amplitude --
one body amplitude of long open (closed) strings. This is 
typical structure of DLCQ partition function. We examined that 
the tadpole cancellation condition in our language and derived 
the well-known gauge group $SO(2^{13})$.
}
\end{abstract}
\vfill

%\vskip.3in
hep-th/0005111
\setcounter{footnote}{0}
\renewcommand{\thefootnote}{\arabic{footnote}}
%\pagestyle{empty}
%\newpage

%%%%%%%%%%%%%%%%%%%%%%%%%%%%%%%%%%%%%%%%%%%%%%%%%%%%%%%%%%%%%%%%%%
\section{Introduction}
%%%%%%%%%%%%%%%%%%%%%%%%%%%%%%%%%%%%%%%%%%%%%%%%%%%%%%%%%%%%%%%%%%
String field theory \cite{KAKU-KIKKAWA}--\cite{KUGO-ASAKAWA-TAKAHASHI} 
has been one of the most fundamental  and the most mysterious subjects
in string theory. In the course of the development, it has been
clarifying the gauge interactions among higher excited
states \cite{WITTEN}\cite{HIKKO}, the moduli problem at least 
for the open string \cite{WITTEN}.
Originally it was regarded as the only candidate
to describe the non-perturbative aspects of string theory.

The revolutionary developments in these years, however,
the new ideas such as D-brane or M-theory turned out to play
more fundamental r\^ole.  One of the shortcoming 
of string field theory may be
that it does not has direct means to describe D-branes dynamics 
although there were some attempts \cite{HATA-HASHIMOTO}.
In this respect, people pay more interests in the 
alternative approaches such as the matrix models
\cite{BFSS}--\cite{MATRIX-REVIEWS} where D-brane itself
becomes the dynamical variable. 

Some years ago, a novel approach \cite{MOTL}--\cite{REVIEWS} to string 
field theory
was evolved from the matrix model view point. In the infrared, the
theory is described as a conformal field theory on the 
symmetric product $S^{n}M=M^{\otimes n}/S_n$.
The orbifold singularities of the target are described as 
the twisted sectors. Excitations belonging to each twisted
sectors can be physically interpreted as the collections
of the ``long strings'' which are composite of the 
fundamental string variable (``short string'' or ``string bit'').
The theory therefore contains a mechanism of the splitting/joining
interactions of the closed string naturally in its definition.
The partition function is expressed as the exponential
of the one body partition function of the long string \cite{DMVV}.
This is typical structure of the partition function of 
the quantum field theory  in the discrete 
lightcone quantization (DLCQ) \cite{SUSSKIND}
which is not restricted to the string theory.
This is the analog of the fact that the vacuum amplitude
can be expressed as the exponential of the contributions
from the connected Feynmann diagrams in the conventional
quantum field theories. These facts support the idea
that the matrix string theory describes the string field
theory in DLCQ.

In this direction, a steady progress was made. 
For example, the four point amplitudes 
of the string theory was directly calculated 
\cite{ARUTYUNOV-FROLOV}\cite{JMR}
by this method to reproduce Virasoro amplitude.
It is applied to describe the little string theory
\cite{DVV2}\cite{DMMV} 
to reproduce the black-hole entropy formula. 
It was generalized to heterotic matrix strings
\cite{BANKS-MOTL}--\cite{REY} to describe the second
quantized lightcone heterotic string.
Some aspects of the $S_N$ orbifold CFT such
as the modular properties and the fusion rule coefficients
are studied in \cite{BANTAY}.
A lot of the developments are made in the context of the
moduli space. In particular, the instanton sectors
of two dimensional Yang-Mills theory is related to the
nontrivial topology of the matrix string world sheet
\cite{WYNTER}--\cite{BRAX}.

{}From the mathematical viewpoint, it is originated from
the calculation of the elliptic genera \cite{DMVV}\cite{KAWAI}
and has a direct relation  with G\"otsche's formula
for Hilbert scheme of points \cite{NAKAJIMA} 
and generalized Kac-Moody algebras.

In this paper, we study the open string version of
the matrix string theory.  The motivation of this subject
should be obvious since we can not escape from dealing with
D-branes in  the matrix strings.
We use the explicit calculation based on
the boundary conformal field theory on the orbifold
\cite{CALLAN}--\cite{BPPZ} and give the some of the explicit
analysis which should be made in BCFT.
The new material is the appearance of the long open (closed)
strings in the twisted sector of the open string\footnote{
While we are finishing this manuscript, we noticed the work
by Johnson \cite{JOHNSON}\cite{JOHNSON2} where the notion of 
the long open strings
as the twisted sectors was already mentioned. His strategy
is to split the closed string amplitude into a product of
the open string amplitude \cite{HORAVA}.}.
We give the classification theorem of all the possible form
of such twisted sectors.
We calculate the partition function
for each twisted boundary conditions and show that
it can be reducible to the amplitude  of the one long
string. An interesting feature is the world sheet topology
of the short string is in general different from
that of the long string.
We develop also the boundary state formalism
and reproduced the amplitude. 
If we sum up all possible boundary conditions,
the partition function can be written as
the exponential of the sum of the long string
partition functions.  This is the typical
form of the partition function in the discrete
lightcone gauge. Finally we confirmed that
the dilaton tadpole cancellation occurs when the
gauge group is famous $SO(2^{13})$ for the bosonic string.

Let us explain the organization of this paper.
We put the main claims at the beginning of each section.
One may first read these parts and skip the
detailed explanation or the proof until it becomes necessary.

In section 2, we give a review of the basic structure of the 
orbifold theory on the symmetric product.
We describe it in detail since some of the explicit 
calculations become essential later.
We emphasize the aspect that it can be formulated as 
the conventional non-abelian orbifold theory.
Namely for the torus amplitude, the consistency conditions
for the twisted boundary condition contain all the information
necessary to reproduce its characteristic feature of 
the string field theory. We also give a review of
the discrete lightcone gauge and derived the typical
form of its partition function.

In section 3, we investigate the constraint on the 
twisted boundary conditions for annulus/M\"obius strip/Klein bottle 
amplitudes and relate it to the various boundary/cross-cap states
\cite{CALLAN}--\cite{DI VECCHIA}.
Open string twisted sectors were discussed in literature
\cite{HARVEY-MINAHAN}--\cite{PRADISI-SAGNOTTI}
mainly for the abelian case. For non-abelian case, we need some
extra care because of the non-commutativity. Because the
open string twisted sector is the main object 
in this paper, we describe it in detail.
The content of this section 
is generic and can be applied to arbitrary non-abelian
orbifold models.

In section 4, we exactly solve the constraints for the symmetric product
orbifold.  The solution for the Klein bottle amplitude is
similar to the torus case.
In the Annulus and M\"obius strip cases,
there are some extra series of the solutions  which will be interpreted
as the contributions of the closed string sector.
The content of this section is mathematical but 
is one of our main claim in this article.

In section 5, we calculate the one loop amplitude
in two ways. First we do it by using the explicit
operator formalism for the flat background.
The calculation itself is technically similar to that of section 2.
Second we calculate it by using multiple cover of the
world sheet. 
The abstract combinatorial solutions in section 4
are translated into the  form of the physically clear
interpretation as the long string amplitudes.
One interesting feature is the appearance of four
sectors for the long string in {\em each} of the annulus
and M\"obius string amplitude for the short string.
Namely, the topology of the world sheet
seen from the long string is in general different from
that for the short string as we already mentioned.

In section 6, we give the explicit form of the
boundary/cross-cap states for the arbitrary twisted sectors.
We argue that the boundary state for the short string
can be classified into three types.  The first one
is the conventional boundary state for the long string.
The second one turns out to be the {\em cross-cap} state
for the long string.  This is one of the origin of the topology 
change.  The third one describes
the connection of the two short strings. It encodes
the nature of the string field theory of the orbifold theory.
The cross-cap states for the short string have the similar classification.
We calculate the inner product between them in order to 
examine the ``modular invariance'' or the tadpole condition
in the next section.

In section 7, we first prove that each of three
open string sectors can be expressed as the
exponential of the one-body amplitudes of the 
long strings of the various scale.  This is quite
natural as the DLCQ partition function.
The annulus amplitude for the short string contains
all types of the amplitudes for the long string.
One interesting aspect is that the torus amplitude
which is contained there has the complex (but discrete)
moduli parameter while the original annulus has only
imaginary part.

In section 8, we examine the tadpole cancellation
of the bosonic string in our context.  We use only
the annulus amplitude (for short string)
to derive the tadpole condition. In this
case there are cancellations among the massless parts
of the boundary states (of the short string) alone. 
Since one-string partition function is exponentiated,
one may reduce the tadpole condition
to each one body problem for the long string.
In this form, one can immediately reproduce
the famous relation such as $SO(2^{13})$.

%\newpage
%%%%%%%%%%%%%%%%%%%%%%%%%%%%%%%%%%%%%%%%%%%%%%%%%%%%%%%%%%%%%%%%%%
\section{Review of closed string on symmetric product}
%%%%%%%%%%%%%%%%%%%%%%%%%%%%%%%%%%%%%%%%%%%%%%%%%%%%%%%%%%%%%%%%%%

\subsection{Orbifold CFT on symmetric product}
Let ${\bf X}^I(\tau, \sigma)$ (${\bf X}^I\in M$,
$I=1,\cdots,N$) be the bosonic
coordinates which define the string embedding on
symmetric product $S^N M$. 
The twisted sectors $\cH_h$ ($h\in S_N$) are defined by
the boundary condition of ${\bf X}$,
\begin{equation}
{\bf X}(\sigma_0,\sigma_1+2\pi)= 
h\cdot{\bf X}(\sigma_0,\sigma_1),
\end{equation}
where
$(h\cdot \bX)^I\equiv\sum_J h^{IJ}\bX^J$.

The modular invariant partition function on torus is,
\begin{eqnarray}
% Z_N(\tau,\bar\tau)&=&\frac{1}{|S_N|}\sum_{
%g,h\in S_N  gh=hg}
%\chi_{h,g}(\tau,\bar\tau)\label{e_torus}\\
%\chi_{h,g}(\tau,\bar\tau) & = & \mbox{Tr}_{\cH_h}(g\cdot \be{
%\tau L_0-\bar\tau \bar{L_0}}),\\
%\be{x}&\equiv & e^{2\pi i x}.
%
 Z_N(\tau,\bar\tau)&=&\frac{1}{|S_N|}\sum_{
{\scriptsize \begin{array}{c}g,h\in S_N\\ gh=hg\end{array}}}
\chi_{h,g}(\tau,\bar\tau)\label{e_torus}\\
\chi_{h,g}(\tau,\bar\tau) & = & \mbox{Tr}_{\cH_h}(g\cdot \be{
\tau L_0-\bar\tau \bar{L_0}}),\\
\be{x}&\equiv & e^{2\pi i x}.
\end{eqnarray}
The summation in $g\in S_N$ is needed to define a projection
onto the $S_N$ invariant subspace. 
The constraint 
\begin{equation}
g\cdot h=h\cdot g 
\label{e_torus_constraint}
\end{equation}
in (\ref{e_torus}) is the consistency condition
of the path integral to assure that the twists
in time and space directions commute.

In the non-abelian orbifold, only the conjugacy class
of $h$ has the invariant meaning since 
$g\cdot \bX\in \cH_{ghg^{-1}}$ if $\bX\in \cH_{h}$.
The summation in (\ref{e_torus}) over $h$ is then replaced by
the summation over the conjugacy class $\cC_i$ of $S_N$.
For a particular element $h\in \cC_i$, the solutions
of (\ref{e_torus_constraint}) are the elements
of the centralizer group $\cN_i$.
With the relation $|S_N|=|\cC_i||\cN_i|$,
(\ref{e_torus_constraint}) can be rewritten as,
\begin{equation}\label{e_torus2}
 Z_N(\tau,\bar\tau)=\sum_{i}
\frac{1}{|\cN_i|}\sum_{g\in \cN_i}
\chi_{h,g}(\tau,\bar\tau),
\qquad h\in \cC_i
\end{equation}
This formula is the generic expression for the arbitrary 
non-abelian orbifold.

For the permutation group $S_N$, the conjugacy group is
labeled by the partition of $N$,
since any group element can be written as a product
of elementary cycles $(n)$ of length $n$,
\begin{equation}\label{e_conjugacy_class}
 [h]=(1)^{N_1}(2)^{N_2}\cdots (k)^{N_k},
\quad
\sum_{n>0} n N_n = N.
\end{equation}
The centralizer of such an element is a semi-direct
product of factors $S_{N_n}$ and $\bZ_n$,
\begin{equation}\label{e_centralizer}
 g\in\cN_h = S_{N_1}\times (S_{N_2}\ltimes \bZ_2^{N_2})\times\cdots
 (S_{N_k}\ltimes \bZ_k^{N_k}).
\end{equation}
The factors $S_{N_n}$ permute the $N_n$ cycles $(n)$, 
while the factors $\bZ_n$ rotate each cycle $(n)$.
The order of the centralizer group is,
\begin{equation}
 |\cN_h| = \sum_{n=1}^k n^{N_n}\ N_n!
\end{equation}

The physical interpretation of these factors are well-known.
Let us first consider the case where $h$ is the element of
the cyclic permutation of $N$ elements,
\begin{equation}
 h=\cycl_N  : \quad \bX^I \rightarrow \bX^{I+1},
\quad I \in \left\{ 0,1,\ldots,N-1\right\}.
\end{equation}
Here superscript $I$ is defined by mod $N$.
The twist by $h$ then gives the boundary condition,
\begin{equation}
 \bX^I(\sigma_0, \sigma_1+2\pi)=\bX^{I+1}(\sigma_0, \sigma_1).
\end{equation}
It means that $N$ short closed strings $\bX^I$ are connected
with each other to form one {\em long} string of length $n$.
For the general situation (\ref{e_conjugacy_class}), 
we will have $N_n$ long strings of length $n$ for $n=1,\ldots,k$.
The short strings that form a long string are
sometimes called {\em ``string bits''}.

In this language, the element
of the centralizer group (\ref{e_centralizer})
has a clear interpretation.  $\bZ_n$ factors are
the rotations of the string bits that constitute a
long string of length $n$. $S_{N_n}$ then reshuffle the
long strings of the same length $n$ as a whole
(figure 1).

%%%%%%%%
 \begin{figure}[ht]
  \centerline{\epsfxsize=8cm \epsfbox{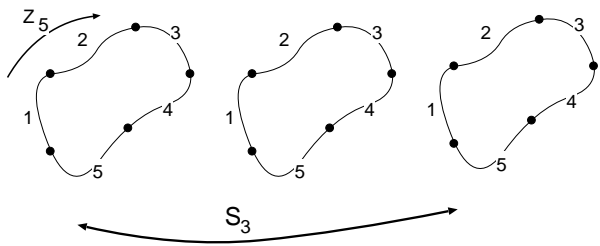}}
  \vskip 3mm
  \caption{Long strings and the action of the centralizer group}
 \end{figure}
%%%%%%%%

\subsection{Partition function}
The long string interpretation can be established
further by the calculation of the partition function.
In this subsection, we would like to give somewhat more
explicit computation compared with the literature
for the preparation for later sections.
To make our argument clear, we first restrict the situation
where target space is flat $M=\bR^1$. 
We will then give a generic argument
for the arbitrary target space.

\subsubsection{Irreducible Diagram}
In the calculation of  the partition function 
(\ref{e_torus2}),
we need to divide $N$ free fields into  small subgroups.
Each subgroup consists of the collection of free fields which
are mixed up by the action of two twists $h$  and $g$.
It is obvious that there are minimal sets of free field
which can not divided into subgroups.
We call such a set of free fields  as ``irreducible
set''.  The partition function $\chi_{h,g}$ is identified
as the product of the contributions of
each irreducible sets.

In the torus case, it is known that such irreducible
sets can be always reduced to $nm$ free fields
where $h$ and $g$ act as,
\begin{eqnarray}\label{e_hg}
 h & = & diag(\cycl_n,\cdots,\cycl_n)\nn\\
 g & = & diag(\cycl_n^{p_1},\cdots,\cycl_n^{p_m})\cdot
 \cyclb_m\,\,,
\end{eqnarray}
where $\cyclb_m$ acts on $\bX^{I,J}$ as
$(\cyclb_m\cdot \bX)^{I,J}=\bX^{I,J+1}$ and
$\cycl_n$ acts as $(\cycl_n\cdot \bX)^{I,J}=\bX^{I+1,J}$.
We label $nm$ free fields as $\bX^{I,J}$ ($I \in(0,1,\cdots,n-1)$
and $J\in(0,1,\cdots,m-1)$). $I$ (resp. $J$) is defined modulo
$n$ (resp. $m$). 
%$\cycl_n$ acts on the first index $I$, and
%$\cycl_m$ acts on the second $J$. 
$p_J$'s represent the rotations each long strings and
take their values in $0,1,\cdots, n-1$.

The action of $h$ and $g$
on $\bX$ in the component is given as,
\begin{eqnarray}
 (h\cdot \bX)^{I,J} & = & \bX^{I+1,J}\nn\\
 (g\cdot \bX)^{I,J} & = & \bX^{I+p_{J},J+1}
\end{eqnarray}
To make $h$ action diagonal, we introduce the
discrete Fourier transformation with respect to $I$,
\begin{equation}\label{e_DF}
 \widetilde{\bX}^{a,J}\equiv  \frac{1}{\sqrt{n}}\sum_{I=0}^{n-1} 
 \be{-\frac{a I}{n}}\bX^{I,J},
\quad a\in (0,1,\cdots,n-1)
\end{equation}
In this basis, the actions of $h,g$ are modified to,
\begin{eqnarray}\label{e_hgX}
 (h\cdot \widetilde\bX)^{a,J} & = & \be{\frac{a}{n}}\widetilde\bX^{a,J}\nn\\
 (g\cdot \widetilde\bX)^{a,J} & = & \be{\frac{ap_{J}}{n}}\widetilde\bX^{a,J+1}
\end{eqnarray}

Since $h$ action is diagonalized, the periodicity
of $\tbX$ becomes well-defined,
\begin{equation}
 \tbX^{a,J}(\sigma_0,\sigma_1+2\pi)=
 \be{\frac{a}{n}}\tbX^{a,J}(\sigma_0,\sigma_1).
\end{equation}
We have the mode expansion of $\widetilde\bX^{a,J}$ as
($z=e^{i(\sigma_1+i\sigma_0)}$),
\begin{equation}\label{e_ME}
 \widetilde\bX^{a,J}=\alpha_0^J\delta_{a,0}\sigma_0
+i \sum_{r\in\bZ}\left(
\frac{1}{r-a/n}\alpha^{a,J}_{r-a/n}z^{-r+a/n}+
\frac{1}{r+a/n}\tilde\alpha^{a,J}_{r+a/n}\bar{z}^{-r-a/n}
\right)
\end{equation}
with the commutation relation,
\begin{equation}
\left[\alpha^{a,I}_{r-a/n},\alpha^{b,J}_{s-b/n}\right]=
(r-a/n)\delta_{I,J}\delta_{r+s}\delta_{a+b}\,\,.
\end{equation}

In order to diagonalize $g$ action, we combine the oscillators
further as,
\begin{equation}
 \bY^{a}\equiv \sum_{J=0}^{m-1}
 C_J \tbX^{a,J}.
\end{equation}
Eigenstate equation $g\cdot \bY^a=\mu \bY^a$ gives
a relation between the neighboring coefficients,
\begin{equation}
 \mu C_{J+1}= \be{\frac{ap_{J}}{n}}C_J.
\end{equation}
The periodicity condition $C_{m}=C_0$ implies,
\begin{equation}
 \mu^m = \be{\frac{ap}{n}},\quad
 p\equiv\sum_J p_{J}
\end{equation}
The eigenvalues for $g$ action are evaluated as,
\begin{equation}
 \mu^{a,b} = \be{\frac{b}{m}+\frac{ap}{nm}},
\quad a=0,1,\cdots,n-1,\quad b=0,1,\cdots,m-1.
\end{equation}
In other word, the action of $g$ is diagonalized  as
\begin{equation}
 (g\, \alpha_{r-a/n})^{a,b} = \mu^{a,b} \alpha_{r-a/n}^{a,b},
\end{equation}
where $\alpha^{a,b}$ are the linear combinations of $\alpha^{a,J}$.

The (chiral) oscillator part of the partition function
can be now evaluated by using mode expansion and
eigenvalues of $g$,
\begin{eqnarray}\label{e_comb}
 &&\prod_{r=1}^\infty \prod_{a=0}^{n-1}\prod_{b=0}^{m-1}
 \left(1-\be{\frac{b}{m}+\frac{ap}{nm}}\be{(r-a/n)\tau}\right)^{-1}\nn\\
 && = \prod_{r=1}^\infty \prod_{a=0}^{n-1}
 \left(1-\be{\frac{ap}{n}}\be{m\tau(r-a/n)}\right)^{-1}\nn\\
 && = \prod_{r=1}^\infty \prod_{a=0}^{n-1}
 \left(1-\be{-p(r-\frac{a}{n})}\be{m\tau(r-a/n)}\right)^{-1}\nn\\
 &&= \prod_{r=1}^\infty 
 \left( 1- \be{\left(\frac{m\tau-p}{n}\right)r}\right)^{-1}
\end{eqnarray}
Thus at least the chiral part of the partition function
from $nm$ string bits are knitted together to give
the partition function of one long string with modified
moduli parameter 
\begin{equation}\label{e_long_string_moduli}
 \frac{m\tau-p}{n}\equiv \tau_{n,m,p}\,\,.
\end{equation}

As for the momentum integration,
only the $a=0$ component have zero mode.
In the remaining $m$ momentum, only $b=0$ part gives
non-trivial inner product when $g$ is inserted.
We conclude that zero-mode contribution comes from
integration of single momentum.
One subtlety is how to fix the normalization constant.  
While changing the variable from the short string to the long one,
Virasoro generators should be modified to
$L_0\rightarrow L_0/n$ \cite{DVV}. 
Next since we have $m$ of such
long strings moving coherently, we have to multiply it
$m$. The kinetic term is thus modified to $\frac{mp^2}{2n}$.
By integrating out $p$, we get
\begin{equation}
1/\sqrt{\mbox{Im} (\tau_{n,m,p})}\,\,.
\end{equation}

By combining the contributions from the anti-chiral part and momentum
integration, the partition function of $nm$ free fields becomes
\begin{equation}\label{e_Znmp}
Z(n,m,\left\{p \right\}|\tau,\bar\tau) =
Z_1\left(\tau_{n,m,p},\bar\tau_{n,m,p}\right)\,\,.
\end{equation}
$Z_1(\tau,\bar\tau)$ is the standard partition function
for one free boson,
\begin{equation}
 Z_1(\tau,\bar\tau)=
\frac{1}{\sqrt{\mbox{Im}\tau}}(\eta(\tau)\bar\eta(\bar\tau))^{-1}.
\end{equation}
As we see in the next paragraph, (\ref{e_Znmp}) is
the generic feature of the symmetric space orbifold.
It may be physically interpreted as a kind of the renormalization.
Namely path integral over the short string variables are replaced by
that of the long string. The two formula
coincides if we replace the moduli parameter.

%A physical interpretation of (\ref{e_Znmp}) is that
%the partition function of $nm$ string bits are combined
%into the partition function of one long string
%with the long string moduli $\frac{m\tau+p}{n}$.

\subsubsection{Generalization of the target space}
\label{s_general_target}

Our consideration in the previous subsection essentially
depends on the flatness of the target space.
We first review the general arguments on  the
arbitrary target space by using the path integral method.

Let $M$ be a general manifold where a string can live
and consider the path integral
on $\bX^{I,J}\in M$ which satisfies
\begin{equation}
 \bX^{I,J}(w+1) = \bX^{I+1,J}(w)\,\,,\qquad
 \bX^{I,J}(w+\tau) = \bX^{I+p_J,J+1}(w)\,\,,
\end{equation}
for each irreducible set $h,g$ as in the previous subsection.
Here $\bX$ is defined on a small parallelogram with period 
$(1,\tau)$. 

These fields $\bX(w)$ can be rewritten from one field $\cX(w)\in M$
by the identification,
\begin{equation}
 \bX^{I,J}(w)  =  \cX\left(w+J\tau+I-P_J\right)\,\,,\quad
P_J  =  \sum_{\ell=0}^{J-1}p_{\ell}\,\,.
\end{equation}
$\cX$ has the following periodicity derived from 
$\bX^{I+n,J}=\bX^{I,J+m}=\bX^{I,J}$,
\begin{equation}
 \cX(w+n) = \cX(w)\,\,,\qquad
 \cX(w+m\tau-p) = \cX(w)\,\,.
\end{equation}
Namely, $\cX(w)$ is defined on $nm$ combinations
of the parallelogram with period $(n,m\tau-p)$.
The moduli of the bigger torus is given by $\tau_{n,m,p}$.

Path integral over variable $\bX$ should be the same 
as that of $\cX$. It implies the generic rule for 
the arbitrary target space,
\begin{equation}
 Z_{h,g}^{\bX}(\tau,\bar\tau) =
Z^{\cX}(\tau_{n,m,p},\bar\tau_{n,m,p}).
\end{equation}

This type of the proof may look too abstract.
One may give more concrete reasoning 
when the target space
is the orbifold $M=T/\Gamma$ where $T$ is a flat space
and $\Gamma$ is a discrete group which may be non-abelian.
The total target space becomes
$T^N/S_N \ltimes \Gamma^N$.
The partition function is written as
\begin{equation}
 Z_N = \frac{1}{|S_N| \cdot |\Gamma|^N}
 \sum_{
\scriptsize
\begin{array}{c}
h,g\in S_N\\
hg=gh
\end{array}}
 \sum_{
\scriptsize
\begin{array}{c}
\left\{\alpha\right\},\left\{\beta\right\}\in \Gamma\\
\left\{\alpha\right\}\cdot \left\{\beta\right\}^h
=\left\{\beta\right\}\cdot \left\{\alpha\right\}^g
\end{array}}\,\,
\mbox{Tr}_{\cH_{h,\left\{\alpha\right\}}}
\left(g\,\beta\, q^{L_0}\bar{q}^{\bar{L}_0}\right).
\end{equation}
Here $\left\{\alpha\right\},\left\{\beta\right\}$
define the boundary condition for $\bX$ for fixed $h,g\in S_N$
as,
\begin{equation}
  \bX^{i}(w+1) = \alpha^{i}\cdot\bX^{h(i)}(w)\,\,,\qquad
 \bX^{i}(w+\tau) = \beta^{i}\cdot\bX^{g(i)}(w)\,\,.
\end{equation}
The condition $\left\{\alpha\right\}\cdot \left\{\beta\right\}^h
=\left\{\beta\right\}\cdot \left\{\alpha\right\}^g$ in the summation
signifies the ``integrability condition'' of these boundary 
conditions,
\begin{equation}
 \alpha^{i}\beta^{h(i)} = \beta^i\alpha^{g(i)}\,\,,
\qquad i=1,2,\cdots,N\,\,.
\end{equation}

%Let us consider the situation where the target space
%is given by $M=T_d/\Gamma$ where $\Gamma$ is a discrete
%group which acts on $d$-torus $T^d$.  For such model,
%the path integral over $\bX$ should be taken over various
%twisted boundary conditions.  

For irreducible variables with $h,g$ given as (\ref{e_hg}),
we replace the index $i$ to a pair $I,J$ and rewrite
the boundary conditions and consistency conditions
\begin{eqnarray}
&& \bX^{I,J}(w+1) = \alpha^{I,J}\cdot\bX^{I+1,J}(w)\,\,,\qquad
 \bX^{I,J}(w+\tau) = \beta^{I,J}\cdot\bX^{I+p_J,J+1}(w)\,\,,\nn\\
&& \beta^{I,J}\cdot\alpha^{I+p_J,J+1} = \alpha^{I,J}\cdot\beta^{I+1,J}.
\label{e_integrability}
\end{eqnarray}
With this type of the constraint, one can always find
a unique set $\gamma^{I,J}\in \Gamma$ which satisfies,
\begin{equation}
 \gamma^{0,0}=1\,,\quad
 \alpha^{I,J}=\gamma^{I,J}(\gamma^{I+1,J})^{-1}\,,\quad
 \beta^{I,J}=\gamma^{I,J}(\gamma^{I+p_J,J+1})^{-1}\,.
\end{equation}
With this twist factor, one may introduce $\cX(w)\in T$
as before which is defined on the bigger torus
generated by $(n,m\tau-p)$ and relate it $\bX^{IJ}$ as
\begin{equation}
 \bX^{I,J}(w) = \gamma^{IJ}\cX(w+J\tau+I-P_J)\,\,.
\end{equation}
It is easy to derive that $\bX$ thus defined satisfies
(\ref{e_integrability}).
In terms of $\cX$ one may develop the operator
formalism as before.
Since $\gamma$ factors are just redefinition of the
identification between the variables, 
The partition function with different $\gamma$ 
will produce identical partition function.
It will give a factor $|\Gamma|^{nm-1}$.
The only nontrivial element is the global monodromy factor
$A,B\in\Gamma$ defined by, 
\begin{equation}
 A=\gamma^{n,0},\quad B=\gamma^{0,m}\,\,,
\end{equation}
which defines the boundary condition for $\chi(w)$,
\begin{equation}
 \cX(w+n) = A^{-1}\,\cX(w)\,,\quad
 \cX(w+m\tau-p) = B^{-1}\,\cX(w)\,.
\end{equation}
We thus arrive at the desired relation,
\begin{equation}
 \frac{1}{|\Gamma|^{nm}} 
 \sum_{
\scriptsize
\begin{array}{c}
\left\{\alpha\right\},\left\{\beta\right\}\in \Gamma\\
\left\{\alpha\right\}\cdot \left\{\beta\right\}^h
=\left\{\beta\right\}\cdot \left\{\alpha\right\}^g
\end{array}}\,\,
Z^{\bX}_{h,\left\{\alpha\right\},g,\left\{\beta\right\}}(\tau,\bar\tau)
=
 \frac{1}{|\Gamma|} 
 \sum_{
\scriptsize
\begin{array}{c}
 A,B\in \Gamma\\
 AB=BA
\end{array}
}\,
Z^{\cX}_{A,B}(\tau_{n,m,p},\bar\tau_{n,m,p}).
\end{equation}
The partition function for $\cX$ can be obtained
by the explicit operator formalism and obviously identical to
the single string partition function.
One may conclude that the formula (\ref{e_Znmp})
should hold for any target space by replacing $Z_1$
by the partition function of the single string
in that target space.

\subsubsection{Generating function of the partition function}
\label{s_GF}
In order to evaluate the whole partition function
(\ref{e_torus}), we need to follow some steps.
\begin{enumerate}
 \item As we mentioned in the previous subsection,
       $\chi_{h,g}$ in (\ref{e_torus}) is expressed
       as the product of the contributions from the irreducible
       sets (\ref{e_Znmp}).  We need to specify how many of
       such irreducible sets are contained in given pair $h,g$.
 \item As for $h$, we use the representation of the conjugacy class
(\ref{e_conjugacy_class}) and use the definition (\ref{e_torus2}). 
 \item For each $h$, $g$ is an element of the centralizer group
       (\ref{e_centralizer}). If we restrict the sector
       $h=\cycl_n^{\otimes N_n}$, $g$ can be written as
       $S_{N_n}\ltimes \bZ_n^{N_n}$.
       Elements in $S_{N_n}$ should be again decomposed into
       conjugacy class as 
       $(1)^{M_{n,1}}(2)^{M_{n,2}}(3)^{M_{n,3}}\ldots$
       with the constraint,
\begin{equation}
 \sum_{s_n=1}^\infty s_n M_{n,s_n}=N_n.
\end{equation}
 \item In this decomposition, we have $M_{n,m}$ irreducible
       subsets for each $n,m$. By summation over elements in $\bZ_n$
       (namely $p_\ell$ in (\ref{e_hg})), we get the expression of 
       $\chi_{h,g}$,
       \begin{eqnarray}
	\chi_{h,g} & = & \prod_{n,m=1}^\infty (n^{m} Z(n,m))^{M_{n,m}}\\
	Z(n,m) & = & \frac{1}{n^m}\sum_{\left\{p\right\}} 
	 Z(n,m,\left\{p\right\}|\tau,\bar\tau)=\frac{1}{n}\sum_{p}
	 Z(n,m,p|\tau,\bar\tau).
	 \label{e_Znm}
       \end{eqnarray}
       We used the fact that the partition function (\ref{e_Znmp})
       depends only on the sum of $\left\{p\right\}$.
 \item For each, $(\left\{N\right\},\left\{M\right\})$,
       the weight factor is
\begin{equation}
\frac{1}{ \prod_{n=1}^\infty n^{N_n}N_n!}
\prod_{n=1}^\infty \left(
\frac{N_n!}{\prod_{s_n=1}^\infty \left(
s_n^{M_{n,s_n}}M_{n,s_n}!
\right)}
\right)
\end{equation}
The first term is the order of the centralizer group
in (\ref{e_torus}). The second term is the product of
the number of the conjugacy class for each $\left\{M\right\}$.
The factor $N_n!$ is canceled between these two terms.
The factor $n^{N_n}$ can be absorbed in $n^m$ in (\ref{e_Znm})
since $N_n=\sum_m m M_{n,m}$.

 \item Assembling every term, we have the following combination,
\begin{equation}\label{e_constraint_NM}
 Z_N(\tau,\bar\tau)=
\sum_{\scriptsize
\begin{array}{c}
N_n,M_{n,m}\geq 1\\
\sum n N_n=N\\
\sum m M_{n,m}=N_n
\end{array}
}
\prod_{n,m\geq 1}
\left(\frac{1}{m}Z(n,m|\tau,\bar\tau)\right)^{M_{n,m}}
.
\end{equation}

\item In the generating function of the partition function,
$$Z(\zeta|\tau,\bar\tau)\equiv\sum_{N=0}^\infty \zeta^N Z_N(\tau,\bar\tau),$$
the constraint in (\ref{e_constraint_NM}) can be removed to give
a free summation over $\left\{M\right\}$,
\begin{eqnarray}
Z(\zeta|\tau,\bar\tau) &=&
\prod_{n=1}^\infty \prod_{m=1}^\infty
\sum_{M_{n,m}=1}^\infty \frac{1}{M_{n,m}!}
\left(\frac{1}{m}Z(n,m|\tau,\bar\tau) \zeta^{n m}\right)^{M_{n,m}}
\nn\\
&=&\prod_{n=1}^\infty \prod_{m=1}^\infty
\exp\left(\frac{1}{m}Z(n,m|\tau,\bar\tau) \zeta^{n m}
\right)\nn\\
&=&
\exp\left(\sum_{n=1}^\infty \sum_{m=1}^\infty
\frac{1}{n m}\sum_{p=0}^{n-1}Z(n,m,p|\tau,\bar\tau) \zeta^{n m}
\right)\nn\\
& = & \exp\left(
\sum_{N=1}^\infty \zeta^N
\cV_N\cdot Z_1(\tau,\bar\tau)
\right),\\
\cV_N\cdot f(\tau,\bar\tau) &\equiv &
\frac{1}{N}\sum_{\scriptsize \begin{array}{c}
 a,d=1,\cdots,N\\b=0,\cdots,d-1\\ad=N
		  \end{array}}
f\left(\frac{a\tau+b}{d},\frac{a\bar\tau+b}{d}\right)
\label{e_Hecke}
\end{eqnarray}
The operator $\cV_N$ which appeared in the final expression is
called Hecke operator.\cite{SERRE}\cite{APOSTOL} 
It maps a modular form to another one
with the same weight and it typically appears in this type 
of calculation.
\end{enumerate}

We may summarize the computation in this subsection into
a theorem,

\vskip 5mm
\noindent{\bf Theorem 1:}
{\em
The generating function of the partition functions
(\ref{e_torus})
are given in the form,
\begin{equation}\label{e_exponential_formula}
 \sum_{N=0}^\infty \zeta^N Z_N(\tau,\bar\tau)
 =e^{\cF(\zeta|\tau,\bar\tau)},\quad
 \cF(\zeta|\tau,\bar\tau)=\sum_{N=1}^\infty \zeta^N
\cV_N\cdot Z_1(\tau,\bar\tau).
\end{equation}
$\cF$ may be regarded as the free energy \cite{GRIGNANI-SEMENOFF}
which consists of
the contributions from the irreducible diagrams.
The partition function $\cV_N Z_1$ for each irreducible
diagram (which is the contribution from $N$ string bits)
is given by applying Hecke operator (\ref{e_Hecke})
to the modular invariant partition function 
for the single string. It is physically interpreted
as the contribution of a single ``long'' string.
}

%%%%%%%%%%%%%%%%%%%%%%%%%%%%%%%%%%%%%%%%%%%%%%%%%%%%%%%%%%%%%%%%%%
\subsection{Discrete lightcone quantization}\label{DLCQ_review}
%%%%%%%%%%%%%%%%%%%%%%%%%%%%%%%%%%%%%%%%%%%%%%%%%%%%%%%%%%%%%%%%%%
The physical interpretation of theorem 1
as a string field theory can be established by introducing
the notion of the discrete lightcone quantization (DLCQ)
\cite{SUSSKIND}
(see also \cite{REVIEWS} for a review). 

Consider first the usual quantum field theory (free boson theory)
in $d$-dimension. We take $x^+$ as the time variable and
write it as $\tau$.
Lagrangian of the system is given by,
\begin{equation}
 S=\int d^d x (\partial_\tau \Phi \partial_- \Phi
+\frac{1}{2}\sum_{i=2}^{d-1}\partial_i\Phi \partial_i \Phi)
\end{equation}
In DLCQ, the light-like direction is compactified by $S^1$ of radius
$R$,
\begin{equation}
 x^- \sim x^- + 2\pi R\,\,.
\end{equation}
Consequently the momentum along that direction is quantized
\begin{equation}
 p^+ = N/R\,\,.
\end{equation}
We expand the (correctly normalized) wave functions along the transverse
coordinates by the eigenfunction of Hamiltonian for the transverse
degree of freedom,
\begin{equation}
 H^{t}\psi_a(\bx) = \lambda_a \psi_a(\bx),
\qquad
a\in\Lambda
\end{equation}
where $\bx$ is the transverse coordinates and 
$\Lambda$ is a set which labels the eigenfunctions.
We introduce the partition function for 
the transverse degree of freedom,
\begin{equation}
 Z_1(\tau) = \sum_{a\in\Lambda} \be{\tau \lambda_a}.
\end{equation}

Fourier transformation along $x^-$ gives the mode expansion
of $\Phi$,
\begin{equation}
 \Phi(x) = \sum_{n=0}^\infty\left(
 \Phi_{n}^a e^{2\pi i nx^-/R}\psi^*_a(\bx)
+ \Phi^{a\dagger}_n(\bx)e^{-2\pi i nx^-/R}\psi_a(\bx)\right)
\,\,,
\end{equation}
together with the commutation relation,
\begin{equation}
 \left[\Phi_n^a,\Phi^{b\dagger}_m\right] =
 \frac{1}{n}\delta_{n,m} \delta_{a,b}.
\end{equation}
Every oscillator has the index $n$ which describes
the sectors classified by the lightcone momentum.

The Hilbert space $\cH_N$ with definite total lightcone momentum 
$p^+=N/R$ is constructed out of these oscillators as,
\begin{equation}\label{e_LC_HILB}
\cH_N\equiv \left\{\left.
\overbrace{\Phi^{a^{1}_1\dagger}_1
\cdots \Phi^{a^{1}_{N_1}\dagger}_1}^{N_1}
\overbrace{\Phi^{a^{2}_1\dagger}_2
\cdots \Phi^{a^{2}_{N_2}\dagger}_2}^{N_2}
\cdots|0\rangle 
\,\,\right| \,\,
a^{\ell}_s\in \Lambda, \,N_n \geq 0, \,\sum_n nN_n=N.
\right\}\,\,.
\end{equation}
DLCQ partition function is defined as the trace over
such Hilbert space with weight $\zeta^N$,
\begin{equation}
 Z(\zeta,\tau)= \sum_{N=0}^\infty 
\mbox{Tr}_{\cH_N}(\zeta^{p^+} \be{\tau p^-})\,\,,
\end{equation}
where $p^+=N/R$. $p^-$ can be expressed as
$
 p^- = \frac{1}{p^+} H^t
$
from the on-shell condition. More explicitly  we assign
the weight factor 
$
\zeta^{\sum_n nN_n/R} \be{\sum_{n,s} R\tau \lambda_{a^n_s} /n}
$
to the state  in the brace in (\ref{e_LC_HILB}).
In the following, we absorb $R$ by redefinition of $\zeta,\tau$
and put $R=1$.

Since we have summation over $N$, it is clear that the partition 
function can be written as a product, $
 Z(\zeta,\tau) =\prod_{n=1}^\infty Z_n(\zeta,\tau),
$
where $Z_n(\zeta,\tau)$ is the contribution from the
states generated by $\Phi^{a\dagger}_n$ for fixed $n$. 
These sub-factors can be computed as follows,
\begin{eqnarray}
 Z_n(\zeta,\tau) & = & \sum_{\left\{N_a\right\}, a\in\Lambda}
\prod_{a\in\Lambda} \zeta^{nN_a} \be{\tau N_a \lambda_a/n}\nn\\
&=& \prod_{a\in\Lambda} \frac{1}{1-\zeta^n \be{\tau \lambda_a/n}}\nn\\
& = & \exp\left(
\sum_{m=1}^\infty \frac{\zeta^{nm}}{m} 
\sum_{a\in\Lambda} \be{\lambda_a m/n}
\right)\nn\\
& = & \exp\left(\sum_{m=1}^\infty \frac{\zeta^{nm}}{m} 
Z_1(m\tau/n)
\right)\,\,.
\end{eqnarray}
In passing from the second to the third lines, we used the formula
$1/(1-x)=e^{\sum_{n=1}^\infty x^n/n}$.
Assembling these terms, we arrive at the formula which is already
very similar to (\ref{e_exponential_formula}),
\begin{equation}
 Z(\zeta,\tau) = \exp( \sum_{n,m=1}^\infty 
\frac{\zeta^{nm}}{m} Z_1(m\tau/n)
)\,\,.
\end{equation}

In case of the open string field theory in DLCQ, we
expect exactly the same formula if we reinterpret $\Lambda$
as the label for the Fock space of one-body open string 
generated by the transverse oscillators 
\cite{KAKU-KIKKAWA}--\cite{GREEN-SCHWARZ}.
We will prove it in section 7 by using only the combinatorics
of the orbifold theory on the symmetric product.

For the closed string field theory, we have to take care of the
winding mode \cite{SUSSKIND} for $x^-$,
\begin{equation}
 \int d\sigma \partial_\sigma X^-(\sigma) = 2\pi R \nu,
\end{equation}
where $\nu$ is the winding number.
By using the relation $\partial_\sigma X^- \sim 
\partial_\sigma X^i\partial_\sigma X^i$, Susskind argued that
the transverse Virasoro generators must satisfy,
\begin{equation}
 L_0 -\bar{L}_0 = \nu n,
\end{equation}
where $n$ is the eigenvalue for $p^+=n/R$.
In this respect, we need to insert the projection operator
which restrict the partition sum into the states
satisfying $L_0-\bar{L}_0\equiv 0$ mod $n$.
This is achieved by the replacement,
\begin{equation}
 Z(\frac{m\tau}{n},\frac{m\bar\tau}{n})
\rightarrow
\frac{1}{n}
\sum_{p=0}^{n-1} Z(\frac{m\tau-p}{n},\frac{m\bar\tau-p}{n})\,\,.
\end{equation}
It gives exactly (\ref{e_exponential_formula}).

At  this point, we find that 
the partition function for the string theory
in DLCQ coincides exactly to the partition
function of orbifold theory if we identify $M=\bR^{24}$
($M=\bR^8$ for the superstring).
The dictionary between the two picture is first the
length of the long string $n$ should be identified with
the lightcone momentum $n/R$.

Comparing the formula in ordinary LCQ of string theory with
the discretized formula, 
(\ref{e_exponential_formula}),
\begin{equation}
\int \frac{d^2\tau}{\mbox{Im} \tau} Z_1(\tau,\bar\tau)
\rightarrow  \sum_{m=1}^\infty  \frac{1}{mn}\sum_{p=0}^{n-1}
Z_1(\tau_{n,m,p},\bar\tau_{n,m,p})\,\,,
\end{equation}
it is natural to regard the summation over $m,p$ as the
discrete version of the integral over $\tau$. Namely
we have the identification, $\mbox{Re} \tau \rightarrow p/n$,
$\mbox{Im} \tau \rightarrow m/n$ 
(see for example, \cite{GRIGNANI-SEMENOFF}).
%This identification is essential in our discussion
%on the tadpole condition.

We note that the moduli parameter $\tau$ for the short string
becomes rather ``redundant'' variable.  We might say that
we do not need integration over moduli for the
constituent string bits and may simply set $\tau=i$.

%%%%%%%%%%%%%%%%%%%%%%%%%%%%%%%%%%%%%%%%%%%%%%%%%%%%%%%%%%%%%%%%%%
%\newpage
%%%%%%%%%%%%%%%%%%%%%%%%%%%%%%%%%%%%%%%%%%%%%%%%%%%%%%%%%%%%%%%%%%
\section{Open string CFT on non-abelian orbifold}
%%%%%%%%%%%%%%%%%%%%%%%%%%%%%%%%%%%%%%%%%%%%%%%%%%%%%%%%%%%%%%%%%%
In this paper, we aim to extend the analysis in 
the previous section to the open string theory.
For that purpose, we need to find the analog of the consistency condition
(\ref{e_torus_constraint}) for various open string
one-loop diagrams (Klein bottle, Annulus, M\"obius strip).
Although this is an elementary issue, we could not
find a literature where such conditions for non-abelian
orbifold were examined\footnote{
Conditions for the abelian orbifold were studied in
many papers. For the incomplete list, 
see \cite{HARVEY-MINAHAN}--\cite{PRADISI-SAGNOTTI}.} 
in detail. 
We note that an important difference between abelian and non-abelian
situations is that we will have non-trivial open
string twisted sectors \cite{HARVEY-MINAHAN} in non-abelian case.
In the abelian orbifold the twisted sector reduces to the standard 
Dirichlet-Neumann sector. 
On the other hand, in the permutation orbifold case,
there are infinite varieties of open string twisted sectors
where they will be interpreted to give the long open (or closed) strings.

In this section, we give a summary of the consistency
conditions for the generic non-abelian orbifold
$M/\Gamma$ where $\Gamma$ is a non-abelian discrete group.

\subsection{Open string Hilbert space}

The usual boundary conditions for the open string
are the following two,
\begin{equation}
 (\bX_R(\sigma_0,\sigma_1)\pm 
\bX_L(\sigma_0,\sigma_1))|_{\sigma_1=0}=0\,\,,
\end{equation}
where $+$ (resp. $-$) sign  is assigned for
Neumann (resp. Dirichlet) boundary condition if $\bX$
is decomposed as
$\bX  =  \bX_L(z) + \bX_R(\bar z)$.
In orbifold case, 
they are generalized to the twisted reflection 
relation between the left and right movers
\begin{equation}\label{e_twist_bc}
 \left.\left(\bX_R(\sigma_0,\sigma_1)\pm f\cdot \bX_L(\sigma_0,\sigma_1)\right)
\right|_{\sigma_1=0}  =  0,
  \qquad f\in\Gamma\,\,.
\end{equation}
As long as the consistency of the boundary state
which we will examine later,
this condition is consistent for the arbitrary element
$f\in\Gamma$. However, such a general twist leads us to 
the unequal footing for the left and the right movers.
This is the situation which appears in the asymmetric orbifold
which we would not like discuss in this article.
In this sense, we will impose a constraint on $f$,
\begin{equation}\label{e_f2}
 f^2=1\,\,.
\end{equation}
When $f$ is not identity in $\Gamma$, taking Dirichlet type boundary
condition means that the open string endpoint
is fixed at the fixed point of the orbifold action.
On the other hand, the interpretation of Neumann 
boundary condition is not very clear.

The open string Hilbert space is specified by
the boundary twists at two boundaries $\sigma_1=0,\pi$.
Let us write it as $\cH_{f_1,f_2}$ where
$f_1\in \Gamma$ (resp.$f_2$) is the twist at
$\sigma_1=0$ (resp. $\sigma_1=\pi$).
We note that by changing the basis for $\bX$,
the boundary condition reduces to $f_{1,2}=\pm 1$,
namely  Dirichlet or Neumann boundary conditions
for the abelian orbifold.

For the permutation group, the boundary condition 
(\ref{e_twist_bc}) has a clear physical interpretation.
Because of the condition (\ref{e_f2}),
$f$ should belong to the conjugacy class of the form,
$(1)^{n_1}(2)^{n_2}$. If $f=\cycl_1$, it simply means that
the open string bit has Neumann or Dirichlet boundary conditions
at that boundary.
On the other hand, if $f=\cycl_2$ or the permutation
of (12), the twisted boundary condition reads
$$
\left.\bX^1_L\pm\bX^2_R\right|_{\sigma_1=0}=0,\quad 
\left.\bX^2_L\pm\bX^1_R\right|_{\sigma_1=0}=0.
$$
For Dirichlet type condition, it means that 
two open string bits are connected smoothly at that boundary.

{\em In the following sections, we will mainly use Neumann 
boundary condition when $f$ defines the open boundary 
and Dirichlet condition when $f$ describes the connection
of two edges.} This restriction is not essential
in our discussion and can be easily generalized to include the
mixed boundary conditions.

By combining $f_1$ and $f_2$, one can realize many varieties of
the configurations for the open string bits. For example, we
illustrate the situation for a long open string
(figure 2, left), $f_1=(1)(23)(4)$, $f_2=(12)(34)$,
and for a long closed string (figure 2, right), $f_1=(14)(23)$, 
 $f_2=(12)(34)$.

%%%%%%%%
 \begin{figure}[ht]
  \centerline{\epsfxsize=10cm \epsfbox{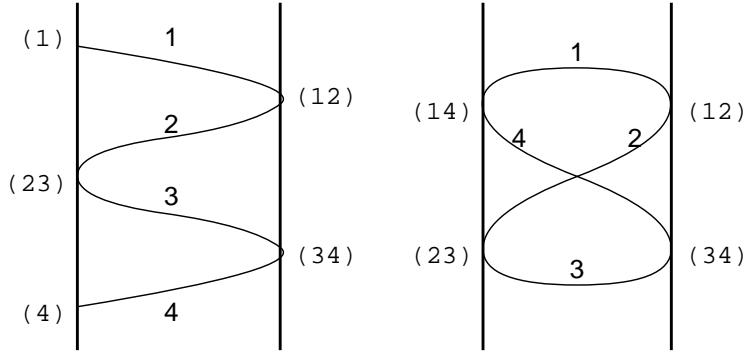}}
  \vskip 3mm
  \caption{Long open (closed) strings}
 \end{figure}
%%%%%%%%
It is clear that by choosing a suitable pair, $(f_1, f_2)$,
one may realize both the long open and closed strings of arbitrary
length.\footnote{For the long closed string, it should be constructed
out of even number of open string bits.}

\subsection{Annulus diagram and boundary state}
The annulus diagram is represented as the trace of
open string Hilbert space $\cH_{f_1,f_2}$ with the projection
operator onto the $\Gamma$ invariant state,
\begin{equation}\label{e_annulus}
\mbox{Tr}_{\cH_{f_1,f_2}}\left(
  g\be{\tau L_0}\right)
\end{equation}
Here and in the following sections, the moduli parameter 
$\tau$ is pure imaginary.

It is straightforward to check that the periodicity along
time direction (twist by $g$) and the twist at the boundary
(\ref{e_twist_bc}) is consistent only if,
\begin{equation}\label{e_annulus_cond}
 g\cdot f_1=f_1\cdot g,\quad g\cdot f_2=f_2\cdot g
\end{equation}

%Unlike the closed string case (\ref{e_torus}), 
%we do not have a strong reason why we should sum
%over $f_1,f_2$ since there are no direct analogue of
%the modular invariance. For the moment, we adopt (\ref{e_annulus})
%since it produces the most possible result
%for the permutation orbifold case as we will see.

For the oscillator representation of
the open string Hilbert space, 
the standard method is to introduce a chiral field
on the double cover (see for example \cite{PRADISI-SAGNOTTI}).
%%%%%%%%
 \begin{figure}[ht]
  \centerline{\epsfxsize=5cm \epsfbox{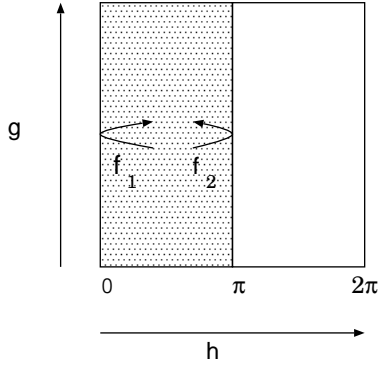}}
  \vskip 3mm
  \caption{Double cover for annulus diagram}
 \end{figure}
%%%%%%%%
For the annulus case depicted in figure 3,
we impose the path integral variable  
$\cX$ to have the twisted boundary condition,
\begin{equation}
 \cX(\sigma_0,\sigma_1+2\pi) = h \cX(\sigma_0,\sigma_1)\qquad
 \cX(\sigma_0+2\pi,\sigma_1) = g \cX(\sigma_0,\sigma_1)
\end{equation}
$g,h\in\Gamma$ should satisfy
$
 g\cdot h= h\cdot g.
$
We will identify it with $\bX$ by (for $0\leq \sigma_1 \leq \pi$)
\begin{equation}
 \bX_L(\sigma_0,\sigma_1)=\cX(\sigma_0,\sigma_1),\quad
 \bX_R(\sigma_0,\sigma_1)=f_2\cdot \cX(\sigma_0,2\pi-\sigma_1).
\end{equation}
With this assignment, $\bX$ satisfies the boundary condition
at $\sigma_1=\pi$ automatically. On the other hand,
the boundary condition at $\sigma_1=0$ requires,
\begin{equation}
 \bX_R(\sigma_0,0)  =  f_2\cdot \cX(\sigma_0,2\pi) = 
 f_2\cdot h\cdot \cX(\sigma_0,0).
\end{equation}
Since the last formula should be same as $f_1\cdot \cX(\sigma_0,0)$,
\begin{equation}\label{e_annulus_twist}
 h=f_2^{-1}\cdot f_1.
\end{equation}
$h$ gives the actual twist in the open string sector
and satisfies $h^g\equiv ghg^{-1}=h$.

The variant of the modular invariance in the open string
case is that the annulus partition function
can be alternatively expressed as the inner product
between the boundary states.
In the orbifold case, the boundary state depends
on two elements $(g,f)$ in the group $\Gamma$.
We will denote it as $\Bket{g}{f}$.

The first element $g$ specifies  which twisted sector
the closed string variable belongs to.
\begin{equation}\label{e_Boundary_state}
\left.(\bbX(\tsigma_0, \tsigma_1+2\pi)-g\cdot \bbX(\tsigma_0,\tsigma_1))
\right|_{\tsigma_0=0} \Bket{g}{f}=0\,\, .
\end{equation}
(We used different notations for embedding functions $\bbX$
and world sheet coordinates $\sigma$ in order to
explicitly shows that we are considering the closed string sector.)
The second element $f$ specifies the twist at the boundary,
\begin{equation}
\left.(\bbX_R(\tsigma_0,\tsigma_1)-f\cdot \bbX_L
(\tsigma_0,\tsigma_1))\right|_{\tsigma_0=0} \Bket{g}{f}=0.
\end{equation}
From the boundary conditions of the open string,
we need to impose the constraints,
\begin{equation}
 \left[g, f\right]=0\,\, ,\quad
 f^2=1\,\, .
\end{equation}
The modular transformation implies the relation,
\begin{equation}
 \mbox{Tr}_{\cH_{f_1,f_2}}\left(g\be{\tau L_0^{open}}\right)
\sim
 \Bbra{g}{f_2}\, \be{-(1/2\tau) (L_0^{closed}+\bar{L}_0^{closed})}
 \Bket{g}{f_1}\,\,.
\end{equation}

%%%%%%%%%%%%%%%%%%%%%%%%%%%%%%%%%%%%%%%%%%%%%%%%%%%%%%%%%%%%%%%%%%
\subsection{M\"obius strip and cross-cap state}
%%%%%%%%%%%%%%%%%%%%%%%%%%%%%%%%%%%%%%%%%%%%%%%%%%%%%%%%%%%%%%%%%%
In M\"obius diagram, we need to calculate the trace such as,
\begin{equation}\label{e_Mobius}
 \mbox{Tr}_{\cH_{f_1,f_2}}\left(\Omega \, g \,
\be{\tau L_0^{open}}\right)\,\, ,
\end{equation}
where $\Omega$ is the open string flip operator.
As in the annulus case, we need to impose
some constraints on $f_1,f_2,g\in\Gamma$
to have a non-vanishing result.

First let us investigate the action of $\Omega g$ on
$\cH_{f_1,f_2}$. It acts on $\bX_L,\bX_R$ as
\begin{eqnarray}
 \bX_L & \longrightarrow & \bY_L (\sigma_0,\sigma_1)
  \equiv g\cdot \bX_R(\sigma_0,\pi-\sigma_1)\nn\\
 \bX_R & \longrightarrow & \bY_R (\sigma_0,\sigma_1)
  \equiv g\cdot \bX_L(\sigma_0,\pi-\sigma_1).
\end{eqnarray}

The boundary condition for the field $\bY$ becomes,
\begin{eqnarray}
 \bY_R(\sigma_0,0) & = & g\bX_L(\sigma_0,\pi)\nn\\
 & = & g\, f_2^{-1} \cdot \bX_R(\sigma_0,\pi)\nn\\
 & = & g\, f_2^{-1} \, g^{-1}\cdot \bY_L(\sigma_0,0)
\end{eqnarray}
In other word, $\bX\in\cH_{f_1,f_2}$ implies
$\Omega\, g\cdot \bX \in \cH_{gf_2^{-1}g^{-1},gf_1^{-1}g^{-1}}$.
In order to have non-vanishing trace (\ref{e_Mobius}),
we need to impose these two Hilbert spaces are the same,
\begin{equation}\label{e_M1}
 f_2 = gf_1^{-1}g^{-1},\quad
 f_1 = gf_2^{-1}g^{-1}.
\end{equation}
If one examines these two conditions carefully, one may notice
that $g,f_1,f_2$ should satisfy further conditions,
\begin{equation}
 \left[g^2,f_1\right]= \left[g^2,f_2\right]=0
\end{equation}
To summarize, the non-vanishing M\"obius strip amplitude
is characterized by $g,f_1\in \Gamma$ with constraints,
\begin{equation}\label{e_M2}
 f_1^2=1,\qquad \left[g^2,f_1\right]=0\,\, .
\end{equation}
$f_2$ is then determined from (\ref{e_M1}).
The actual twist in the open string is given by,
\begin{equation}\label{e_M3}
 h=f_2^{-1}\,f_1= g \, f_1\, g^{-1}\, f_1\,\,,
\end{equation}
which satisfies the condition typical in unorientable cases,
$
 h^g=h^{-1}
$.

%%%%%%%%
 \begin{figure}[ht]
  \centerline{\epsfxsize=4cm \epsfbox{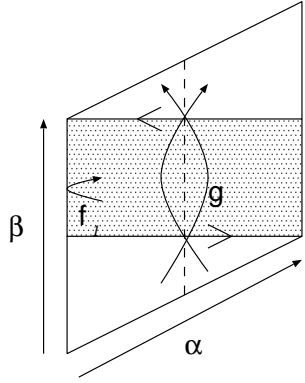}}
  \vskip 3mm
  \caption{Double cover for M\"obius diagram}
 \end{figure}
%%%%%%%%
In the standard double covering of M\"obius strip (figure 4),
these conditions are explained as follows.
We introduce the path integral variable  $\cX$ satisfies 
the twisted boundary condition,
\begin{eqnarray}\label{e_Mobius_chiral}
 \cX(\sigma_0+\pi, \sigma+\pi) &=& \alpha\cdot \cX(\sigma_0,\sigma_1)
\nn\\
 \cX(\sigma_0+2\pi,\sigma) &=& \beta\cdot \cX(\sigma_0,\sigma_1)
\qquad
\alpha,\beta\in\Gamma.
\end{eqnarray}
From this field, we would like to construct the left and right
movers of the open string which satisfy the boundary condition,
\begin{eqnarray}
 \bX_R(\sigma_0,0) & = & f_1\cdot \bX_L(\sigma_0,0)\nn\\
 \bX_R(\sigma_0+\pi,\pi/2) & = & g\cdot \bX_L(\sigma_0,\pi/2),\nn\\
 \bX_L(\sigma_0+\pi,\pi/2)  &=&  g\cdot \bX_R(\sigma_0,\pi/2).
\end{eqnarray}
The boundary conditions in the second and the third lines
are (twisted) cross-cap type conditions.
It is consistent with (\ref{e_Mobius_chiral})
only when $\beta=g^2$.

We use the following  identification between
$\bX$ and $\cX$,
\begin{eqnarray}
 \bX_L(\sigma_0,\sigma_1)&=&\cX(\sigma_0,\sigma_1)\nn\\
 \bX_R(\sigma_0,\sigma_1)&=&f_3\cdot \cX(\sigma_0+\pi,\pi-\sigma_1).
\end{eqnarray}
Boundary condition at $\sigma_1=\pi/2$ implies,
\begin{eqnarray}
 \bX_R(\sigma_0+\pi,\pi/2) & = & f_3\cdot \cX(\sigma_0+2\pi,\pi/2)\nn\\
 & = & f_3\, g^2 \cdot \cX(\sigma_0,\pi/2)\nn\\
 g\cdot \bX_L(\sigma_0,\pi/2) & = & g\cdot \cX(\sigma_0,\pi/2).
\end{eqnarray}
Namely,
\begin{equation}
 f_3=g^{-1}.
\end{equation}
Similarly boundary condition at $\sigma_1=0$ is consistent
with (\ref{e_Mobius_chiral}) only when,
\begin{equation}
 \alpha=g\,f_1
\end{equation}
Periodicity   and the twisted boundary condition
along $\sigma_1=0$ implies,
\begin{equation}
 \left[f_1,\beta\right]=\left[f_1,g^2\right]=0.
\end{equation}

This calculation can be summarized by the introduction of
the cross-cap state $\Ckets{g}{f}$ which satisfies,
\begin{eqnarray}
 (\bbX_L(0,\tsigma_1+\pi)-f\cdot\tbX_R(0,\tsigma_1))\Ckets{g}{f}&=&0
\nn\\
 (\bbX_R(0,\tsigma_1+\pi)-f\cdot\tbX_L(0,\tsigma_1))\Ckets{g}{f}&=&0
\label{e_crosscap}
\end{eqnarray}
The twisted boundary condition of the closed string
is specified by $g$.  However, 
it is clear that these conditions automatically implies that
$\bbX$ belongs to the twisted sector defined by $f^2$,
\begin{equation}
 \bbX(\tsigma_0,\tsigma_1+2\pi) = f^2\cdot \bbX(\tsigma_0,\tsigma_1).
\end{equation}
Namely we always have $g=f^2$.
The modular transformation of  M\"obius strip can 
be now written as
\begin{equation}
 \mbox{Tr}_{\cH_{f_1,f_1^g}}\left(
g\,\be{\tau L_0^{open}}\right)
= \Bbra{g^2}{f_1}\, \be{-(1/8\tau)(L_0^{closed}+\bar{L}_0^{closed})}
\Cket{g}.
\end{equation}

\subsection{Klein bottle}
In the Klein bottle amplitude, we need to evaluate
\begin{equation}
 {\mbox{Tr}}_{\cH_h}\left(\Omega^{closed}\,g\,
 \be{\tau(L_0^{closed}+\bar{L}_0^{closed})}\right)
\end{equation}
As in the M\"obius strip case, 
\begin{equation}
\bX\in \cH_h\rightarrow
\Omega^{closed}g\bX\in \cH_{gh^{-1}g^{-1}}.
\end{equation}
Therefore, to have non-vanishing trace, we need impose
constraint on $g,h$,
\begin{equation}\label{e_KB1}
 h=gh^{-1}g^{-1}\quad\mbox{or}\quad hg=gh^{-1}.
\end{equation}

%%%%%%%%
 \begin{figure}[ht]
  \centerline{\epsfxsize=4cm \epsfbox{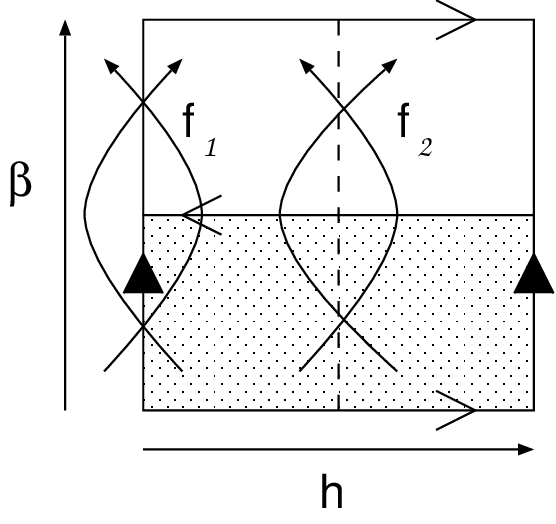}}
  \vskip 3mm
  \caption{Double cover for Klein bottle diagram}
 \end{figure}
%%%%%%%%

In the double covering (figure 5), we define a chiral field
$\cX$ which has twisted boundary condition,
\begin{equation}
 \cX(\sigma_0,\sigma_1+2\pi) = h\cdot\cX(\sigma_0,\sigma_1),
\quad
 \cX(\sigma_0+2\pi,\sigma_1) = \beta\cdot\cX(\sigma_0,\sigma_1).
\end{equation}
$\bX$ satisfies cross-cap type boundary condition at
$\sigma_1=0,\pi$,
\begin{equation}
 \bX_{R,L}(\sigma_0+\pi,0) =f_1 \bX_{L,R}(\sigma_0,0)
\quad
 \bX_{R,L}(\sigma_0+\pi,\pi) =f_2 \bX_{L,R}(\sigma_0,\pi)
\end{equation}
We identify
\begin{equation}
 \bX_L(\sigma_0,\sigma_1)=\cX(\sigma_0,\sigma_1)\quad
 \bX_R(\sigma_0,\sigma_1)=f_2\cdot\cX(\sigma_0-\pi,2\pi-\sigma_1)
\end{equation}
It automatically satisfies boundary condition at $\sigma_1=\pi$.
Boundary condition at $\sigma_1=0$ is satisfied if
\begin{equation}
 h=f_2^{-1}f_1.
\end{equation}
Twist in time direction requires,
\begin{equation}\label{e_KB2}
 \beta=f_1^2=f_2^2.
\end{equation}
If we identify $g=f_1$, two conditions (\ref{e_KB1}) and
(\ref{e_KB2}) are equivalent.
The modular invariance in this case can be written as,
\begin{equation}
 \mbox{Tr}_{\cH_h}\left(g\,\Omega^{closed}\,
 \be{\tau(L_0+\bar{L}_0)}\right)
= \Cbra{f_2}\,\be{-(1/4\tau) (L_0+\bar{L}_0)}\Cket{f_1}
\end{equation}
with $g=f_1$ and $h=f_2^{-1}f_1$.

%%%%%%%%%%%%%%%%%%%%%%%%%%%%%%%%%%%%%%%%%%%%%%%%%%%%%%%%%%%%%%%%%%
%\newpage
%%%%%%%%%%%%%%%%%%%%%%%%%%%%%%%%%%%%%%%%%%%%%%%%%%%%%%%%%%%%%%%%%%
\section{Classification of irreducible boundary conditions}
%%%%%%%%%%%%%%%%%%%%%%%%%%%%%%%%%%%%%%%%%%%%%%%%%%%%%%%%%%%%%%%%%%

In this section, we explicitly solve the constraints of the twists
in the previous section for three diagrams.
We classify all the possible irreducible solutions
together with the free parameters.
Our result in this section is summarized in the following theorem.
\vskip 5mm

\noindent{\bf Theorem 2}\newline
{\em
(i) Klein bottle: irreducible solutions for
(\ref{e_KB1}) are given by,
\begin{equation}\label{e_K}
 h=diag(\overbrace{\cycl_n,\cdots,\cycl_n}^m)\,\,,\quad
 g=diag(\cycl_n^{p_0}\inv_n,\cdots,\cycl_n^{p_{m-1}}\inv_n)\cdot
 \cyclb_m,
\end{equation}
where $\inv_n$ is the inversion permutation 
of $n$ elements $(n-1,n-2,n-3,\cdots,0)$.
$p_i$ ($i\equiv0,\cdots,m-1$ mod $m$) takes their values in 
$0,1,\cdots,n-1$ (mod $n$).
We will refer this solution as (K).
\vskip 2mm
%%%%%%%%%%%%%%%%%%%%%%%%%%%%%%
% Annulus
%%%%%%%%%%%%%%%%%%%%%%%%%%%%%%
\noindent (ii) Annulus: there are three types of the 
irreducible solutions for 
(\ref{e_annulus_cond},\ref{e_annulus_twist})
together with $f_i^2=1$.
\begin{itemize}
 \item ($I_A$):
\begin{eqnarray}\label{e_I_A}
 h & = & diag(\overbrace{\cycl_n,\cdots,\cycl_n}^m)\nn\\
 g & = & diag(\cycl_n^{p_0},\cdots,\cycl_n^{p_{m-1}})\cdot\cyclb_m,\nn\\
 f_1 & = & diag(\cycl_n^{q_0}\inv_n,\cdots,\cycl_n^{q_{m-1}}\inv_n),\nn\\
 f_2 & = & diag(\cycl_n^{q_0+1}\inv_n,\cdots,\cycl_n^{q_{m-1}+1}\inv_n).
\end{eqnarray}
$p_i,q_i$ ($i\equiv0,\cdots,m-1$ mod $m$) takes their values in 
$0,1,\cdots,n-1$ (mod $n$). They should satisfy the constraint,
\begin{equation}\label{e_I_A_local}
 q_\ell -q_{\ell+1} \equiv  2p_\ell,\quad
(\ell=0,\cdots,n-1 \mbox{ mod}\,\,\, n)
\end{equation}
\item ($II_A$): for even $m$, 
with the same $h,g$ as in (\ref{e_I_A}), together
with
\begin{eqnarray}\label{e_II_A}
 f_1 & = & diag(\cycl_n^{q_0}\inv_n,\cdots,\cycl_n^{q_{m-1}}\inv_n)
(\cyclb_{(m)})^{m/2},\nn\\
 f_2 & = & diag(\cycl_n^{q_0+1}\inv_n,\cdots,\cycl_n^{q_{m-1}+1}\inv_n)
(\cyclb_{(m)})^{m/2},
\end{eqnarray}
with $q_{k}=q_{k+m/2}$.
The constraint for $p,q$ are given by
\begin{equation}\label{e_II_A_local}
 q_\ell-q_{\ell+1} \equiv p_\ell+p_{\ell+m/2} \quad
(\mbox{mod}\, \, n)
\end{equation}
$\ell$ in this equation is defined in modulo $m$.
\item ($\widetilde{II}_A$): for even $m$, 
with the same $h,f_1,f_2$ as in ($II_A$), together
with
\begin{equation}\label{e_IIt_A}
 g  =  diag(\cycl_n^{p_0},\cdots,\cycl_n^{p_{m-1}})
diag(\cyclb_{m/2},\cyclb_{m/2}).
\end{equation}
The constraint for $p,q$ are given by
\begin{equation}\label{e_IIt_A_local}
 q_\ell-q_{\ell+1} \equiv p_\ell+p_{\ell+m/2} \quad
(\mbox{mod}\, \, n)
\end{equation}
The index $\ell$ in this equation is defined in modulo $m/2$.
\end{itemize}
\vskip 2mm
%%%%%%%%%%%%%%%%%%%%%%%%%%%%%%
% Mobius
%%%%%%%%%%%%%%%%%%%%%%%%%%%%%%
\noindent (iii) M\"obius strip: there are three types of the 
irreducible solutions for 
(\ref{e_M2},\ref{e_M3}).
\begin{itemize}
 \item ($I_M$):
\begin{eqnarray}\label{e_I_M}
 h & = & diag(\overbrace{\cycl_n,\cdots,\cycl_n}^m)\nn\\
 g & = & diag(\cycl_n^{p_0}\inv_n,\cdots,\cycl_n^{p_{m-1}}\inv_n)
\cdot\cyclb_m,\nn\\
 f_1 & = & diag(\cycl_n^{q_0}\inv_n,\cdots,\cycl_n^{q_{m-1}}\inv_n).
\end{eqnarray}
The constraint for $p,q$ is
\begin{equation}\label{e_I_M_local}
 q_\ell +q_{\ell+1} \equiv  2p_\ell-1,\quad
(\ell=0,\cdots,n-1 \mbox{ mod\,} n)
\end{equation}
\item ($II_M$): for even $m$, 
with the same $h,g$ as in (\ref{e_I_M}), together
with
\begin{equation}\label{e_II_M}
 f_1  =  diag(\cycl_n^{q_0}\inv_n,\cdots,\cycl_n^{q_{m-1}}\inv_n)
(\cyclb_{(m)})^{m/2}.
\end{equation}
The constraint for $p,q$ are given by
\begin{equation}\label{e_II_M_local}
 q_\ell+q_{\ell+1} \equiv p_\ell+p_{\ell+m/2}-1 \quad
(\mbox{mod}\, \, n)
\end{equation}
$\ell$ in this equation is defined in modulo $m$.
\item ($\widetilde{II}_M$): for even $m$, 
with the same $h,f_1$ as in ($II_M$), together
with
\begin{equation}\label{e_IIt_M}
 g  =  diag(\cycl_n^{p_0}\inv_n,\cdots,\cycl_n^{p_{m-1}}\inv_n)
\cdot diag(\cyclb_{m/2},\cyclb_{m/2}).
\end{equation}
The constraint for $p,q$ are given by
\begin{equation}\label{e_IIt_M_local}
 q_\ell+q_{\ell+1} \equiv p_\ell+p_{\ell+m/2}-1 \quad
(\mbox{mod}\, \, n)
\end{equation}
The index $\ell$ in this equation is defined in modulo $m/2$.
\end{itemize}
} %the end of emphasis mode
\vskip 10mm
%%%%%%%%%%%%%%  End of Theorem 2 %%%%%%%%%%%%%%%%%%%%%%%%%

While $\cycl_n^p$ represents the cyclic rotation of
string bits in length $n$ long string,
$\inv_n$ 
flips the orientation of the long string.
Four sectors, $(II_A)$, $(\widetilde{II}_A)$,
$(II_M)$, $(\widetilde{II}_M)$, can be interpreted as
giving the closed string sectors with/without orientation flip
which appears in open string sector.

The rest of this section is devoted to straightforward
but rather lengthy proof of this theorem.
Before we embark on the calculation of
indivisual cases, we first mention a simple lemma
\vskip 2mm

{\em
\noindent{\bf Lemma 1:}
In the irreducible sets,
$h$ should always be the direct product of the cyclic
permutations of the same length $n$.
}
\vskip 2mm
\noindent {\em Proof:} 
We already mentioned that for
each diagram, $h$ and $g$ satisfy,
\begin{eqnarray}\label{e_hga}
 h^g =&  h, & \qquad \mbox{Annulus}\nn\\
 h^{g} =&  h^{-1},&\qquad \mbox{Klein bottle, M\"obius strip},
\end{eqnarray}
where $h^g\equiv ghg^{-1}$.
Let us assume that the conjugacy class of $h$ is given by
the partition $(n)(m)$ with $n\neq m$.
LHS of (\ref{e_hga}) belongs to the same conjugacy class
with each element permuted by $g$. Since $n\neq m$,
$g$ can not mix elements in $(n)$ and $(m)$.

For open string sectors (Annulus and M\"obius), we also
determine $f_i$ ($i=1,2$) with $f_i^2=1$ and $h=f_2^{-1}\, f_1$.
Because of $f_i^2=1$, one may easily derive 
that $h^{f_1}=h^{-1}$.
By repeating our argument on $g$, $f_1$ (and also $f_2$) 
can not mix the elements in $(n)$ and $(m)$. 

It proved that when $h=(n)(m)$ with $n\neq m$, there
are no the irreducible sets. \qed

\subsection{Klein bottle}
{}From lemma 1, we may restrict $h,g$ to the following form,
\begin{equation}
 h=diag(\overbrace{\cycl_n,\cdots,\cycl_n}^m),
\quad
 g=diag(\alpha_0,\cdots,\alpha_{m-1})\cdot\cyclb_m,
\end{equation}
where $\alpha_\ell$ is any permutations of $n$ elements.
$h^g=h^{-1}$ implies, $\alpha_\ell\cycl_n\alpha_\ell^{-1}=\cycl_n^{-1}$.
To derive conditions in (K), it is enough to prove
 the following lemma,
\vskip 2mm
{\em\noindent{\bf Lemma 2:}
The general solution to 
\begin{equation}\label{e_inv}
 g\cycl_ng^{-1} = \cycl_n^{-1},
\end{equation}
is given by $g=\cycl_n^{p}\inv_n$ for $p=0,1,\cdots, n$.
}
\vskip 2mm
\noindent{\em Proof:}
Assume that $g$ maps $(0,1,2,\cdots,n-1)$ to 
$(g_0,g_1,\cdots,g_{n-1})$. $g\cycl_n=\cycl_n^{-1}g$ implies
\begin{equation}
 g(i+1)-g(i)+1\equiv 0 \quad \mbox{mod}\, n.
\end{equation}
General solution to this difference equation is clearly
(\ref{e_inv}).
\qed.

\subsection{Annulus}
Because $\left[h,g\right]=0$, we may write general irreducible
solutions in the following form,
\begin{eqnarray}
 h & = & diag(\overbrace{\cycl_n,\cdots,\cycl_n}^m)\nn\\
 g & = & diag(\cycl_n^{p_0},\cdots,\cycl_n^{p_{m-1}})\cdot G,\nn\\
 f_i & = & diag(\alpha^{(i)}_1 ,\cdots,\alpha^{(i)}_{m-1})\cdot F_{i}.
\end{eqnarray}
where $G$ and $F_{i}$ belong to $S_m$.
The constraints $\left[g,f_i\right]=0$,
$h=f_2^{-1}f_1$ and $f_i^{2}=1$ implies,
\begin{eqnarray}\label{e_FG1}
 \left[G,F_i\right] & = & 0\nn\\
 F_1^2=F_2^2=F_1\,F_2 & = & 1.
\end{eqnarray}
The equation in the second line implies $F_1=F_2$.
We can use Lemma 1  to show that
$G$ is written as a direct product
of the cyclic permutations of the same length 
$(\cyclb_{m/s})^{\otimes s}$
if we use the irreducibility.
From the first equation, $s$ must be either 1 or 2.
If $s=1$, $F_i$  can be either 1 ($I_A$),
\begin{equation}\label{e_IA}
 G=\cyclb_m\,\,\,\qquad F=1_m,
\end{equation}
 or $\cyclb_{m/2}$ if $m$
is even ($II_A$)
\begin{equation}\label{e_IIA}
 G=\cyclb_{m}\,\,\,\qquad 
 F=G^{m/2}=\left( \begin{array}{c c}
    0 & 1_{m/2} \\ 1_{m/2}&0 
	  \end{array}\right)\,\,\,,
\end{equation}
If $s=2$, $G$ and $F_i$ must be the following form ($\widetilde{II}_A$),
\begin{equation}\label{e_IItA}
 G=\left( \begin{array}{c c}
    \cyclb_{m/2}&0 \\ 0 & \cyclb_{m/2}
	  \end{array}\right)\,\,\,, \quad
 F=\left( \begin{array}{c c}
    0 & 1_{m/2} \\ 1_{m/2}&0 
	  \end{array}\right)\,\,\,,
\end{equation}
with even $m$.

\subsubsection{$I_A$}
$f_i^2=1$ and $f_2^{-1}f_1=h$ implies
\begin{equation}
 (\alpha_\ell^{(2)})^{-1} \,
\alpha_\ell^{(1)} = \cycl_n,\qquad
 (\alpha_\ell^{(i)})^2 =1.
\end{equation}
(\ref{e_I_A}) is easily derived by using the following lemma.
(\ref{e_I_A_local}) comes from the constraint $\left[f_i,g\right]=0$.
\vskip 2mm

{\em\noindent{\bf Lemma 3:}
For $f_1,f_2\in S_n$, 
the general solution to 
\begin{equation}
 f_1^2=f_2^2=1,\quad f_2^{-1} f_1=\cycl_n
\end{equation}
is given by
\begin{equation}\label{e_sol_fi}
 f_1=\cycl_n^{p}\inv_n,\quad
 f_2=\cycl_n^{p+1}\inv_n,\quad
 p=0,\cdots,n-1 \quad \mbox{mod}\,\, n.
\end{equation}
}
\noindent{\em Proof:}
If we write $f_2^{-1}f_1=h$ and $f_1=g$, 
\begin{equation}
 h^g=f_1\,f_2 =h^{-1}=\cycl_n^{-1}.
\end{equation}
This is exactly the same condition as Lemma 2.
It permits us to write $f_i$  in the form
(\ref{e_sol_fi}). By using the relation,
\begin{equation}
 \cycl_n^p\inv_n=\inv_n\cycl_n^{-p},
\end{equation}
it is straightforward to prove that $f_i$s thus defined
satisfy $f_i^2=1$.
\qed.

\subsubsection{$II_A$, $\widetilde{II}_A$}
We write $m/2\equiv k$.
The constraints $f_i^2=1$  are written in components,
$\alpha^{(i)}_\ell\alpha^{(i)}_{\ell+k}=1$, namely
$\alpha^{(i)}_{\ell+k}=(\alpha^{(i)})^{-1}_{\ell}$.
By plugging  it into $f_2^{-1}f_1=h$, we get
$$
\alpha_\ell^{(2)}(\alpha_\ell^{(1)})^{-1}=
(\alpha_\ell^{(2)})^{-1}\alpha_\ell^{(1)}=
\cycl_n.
$$ 
From this equation, it is not difficult to prove
$\alpha_\ell^{(i)}\cycl_n\alpha_\ell^{(i)\,-1}=\cycl_n^{-1}$. 
By using lemma 2, we conclude that  $f_i$ must be of the form,
(\ref{e_II_A}). 
Finally constraints (\ref{e_II_A_local},\ref{e_IIt_A_local})
are obtained by imposing $\left[g,f_i\right]=0$.

%%%%%%%%%%%%%%%%%%%%%%%%%%%%%%%%%%%%%%%%%%%%%%%%%%%%%%%%%%%%%%%%%%
\subsection{M\"obius}
%%%%%%%%%%%%%%%%%%%%%%%%%%%%%%%%%%%%%%%%%%%%%%%%%%%%%%%%%%%%%%%%%%

With the help of Klein bottle calculation, 
one may seek the general irreducible
solution in the following form,
\begin{eqnarray}
 h & = & diag(\overbrace{\cycl_n,\cdots,\cycl_n}^m)\nn\\
 g & = & diag(\cycl_n^{p_0}\inv_n,
\cdots,\cycl_n^{p_{m-1}}\inv_n)\cdot G\,\,\,,\nn\\
 f_1 & = & diag(\alpha_0 ,\cdots,\alpha_{m-1})\cdot F_1\,\,,
\end{eqnarray}
with $G,F_1\in S_m$. In this form $h^g=h^{-1}$ is automatically
satisfied. (\ref{e_M2},\ref{e_M3}) lead to
\begin{equation}
 \left[F_1,G\right] =0\,,\quad
 F_1^2=1\,.
\end{equation}
This is exactly the same constraint in the annulus 
(\ref{e_FG1}).
We may use the same solutions, (\ref{e_IA},\ref{e_IIA},\ref{e_IItA})
and call the solutions associated with each of them as
$I_M$, $II_M$, $\widetilde{II}_M$.

\subsubsection{$I_M$}
$f_1^2=1$ implies $\alpha_\ell^2=1$.
Constraint (\ref{e_M3}) gives 
\begin{equation}\label{e_M4}
 \cycl_n^{p_\ell} \inv_n \alpha_\ell^{-1} \cycl_n^{p_\ell} \inv_n
\alpha_{\ell+1} =\cycl_n.
\end{equation}
Since $\alpha_\ell^2=1$ and 
$(\cycl_n^{p_\ell} \inv_n \alpha_\ell^{-1} \cycl_n^{p_\ell}
\inv_n)^2=1$, lemma 3 permits us to write
\begin{equation}
 \alpha_\ell=\cycl_n^{q_\ell}\inv_n\,\,,
\end{equation}
which proved (\ref{e_I_M}). The constraint (\ref{e_I_A_local})
comes from putting this value into (\ref{e_M4}) again.

\subsubsection{$II_M$ and $\widetilde{II}_M$}
In these cases, $f_1$ has the following form ($m=2k$),
\begin{equation}
 f_1=diag(\alpha_0,\cdots,\alpha_{m-1})\cdot (\cyclb_{2k})^{k}.
\end{equation}
$f_1^2=1$ gives $\alpha_{\ell+k}=\alpha_\ell^{-1}$.
We note that $f_2=(f_1)^g$ have the same form.
As in the $II_A$ case, one can show that
$\alpha_\ell$ have the following form,
\begin{equation}
 \alpha_\ell=\cycl_n^{q_\ell}\inv_n\,\,.
\end{equation}
(\ref{e_M3}) gives the same type of constraint in both
$II_M$ and $\widetilde{II}_M$\footnote{
The difference between those two cases is whether the variable
$\ell$ is counted as mod $2k$ ($II_M$) or as mod $k$
($\widetilde{II}_M$).
},
\begin{equation}
 (\cycl_n^{p_\ell}\inv_n)\,\alpha_\ell\,(\cycl_n^{p_{k+\ell}}\inv_n)\,
 \alpha_{\ell+1}^{-1} = \cycl_n\,\,,
\end{equation}
which gives (\ref{e_II_M_local},\ref{e_IIt_M_local}).

The proof of theorem 2 is completed.

%%%%%%%%%%%%%%%%%%%%%%%%%%%%%%%%%%%%%%%%%%%%%%%%%%%%%%%%%%%%%%%%%%
\newpage
%%%%%%%%%%%%%%%%%%%%%%%%%%%%%%%%%%%%%%%%%%%%%%%%%%%%%%%%%%%%%%%%%%
%%%%%%%%%%%%%%%%%%%%%%%%%%%%%%%%%%%%%%%%%%%%%%%%%%%%%%%%%%%%%%%%%%
\section{Partition functions}
%%%%%%%%%%%%%%%%%%%%%%%%%%%%%%%%%%%%%%%%%%%%%%%%%%%%%%%%%%%%%%%%%%

In this section, we explicitly calculate the partition functions
for the irreducible boundary conditions discussed in 
previous section. One interesting feature
is that the sectors for the long strings are in general
different from those of the short strings.
Actually this is clear since we already mentioned
there are the long closed string sectors in the annulus
or M\"obius strip amplitude in the short string.

Since the correspondence looks rather complicated, we summarize 
our result in the following table.

\begin{center}
\begin{tabular}[b]{|c|c|c|c|c|}\hline
 Short string sector& $n$&$m$ & Long string sector& Partition function\\\hline
 KB& $*$ & odd &Klein Bottle & $\cZ^{KB}(\tau_{n,m,0})$\\
   & $*$ & even & Torus& $\cZ^{T}(\tau_{n,m,p},\bar\tau_{n,m,p})$\\\hline
 Annulus: $I_A$ & odd & $*$ & Annulus & $\cZ^{A}(\tau_{n,m,0})$ \\
 & even & $*$ & Annulus+M\"obius &
$\cZ^{A}(\tau_{n,m,0})+\cZ^{M}(\tau_{n,m,0})$ \\
 Annulus: $II_A$ & $*$ & $2\times *$ & Klein Bottle & 
$\cZ^{KB}(\tau_{n,m/2,0})$\\
 Annulus: $\widetilde{II}_A$ & $*$ & $2\times *$ & Torus&
$\cZ^{T}(\tau_{n,m/2,p},\bar\tau_{n,m/2,p})$\\\hline
 M\"obius: $I_M$ & odd & odd & M\"obius &
$\cZ^{M}(\tau_{n,m,0})$\\
  & odd & even & Annulus& $\cZ^{A}(\tau_{n,m,0})$ \\
  & even & even &Annulus+M\"obius
& $\cZ^{A}(\tau_{n,m,0})+\cZ^{M}(\tau_{n,m,0})$\\
  & even & odd & --- &0\\
 M\"obius: $II_M$ & $*$ & $2\times$odd & Torus &
$\cZ^{T}(\tau_{n,m/2,p^*},\bar\tau_{n,m/2,p^*})$ \\
  & $*$ & $2\times$even & Klein Bottle & $\cZ^{KB}(\tau_{n,m/2,0})$\\
 M\"obius: $\widetilde{II}_M$ & $*$ & $2\times$even & Torus &
$\cZ^{T}(\tau_{n,m/2,p},\bar\tau_{n,m/2,p})$\\
  & $*$ & $2\times$odd & Klein Bottle & $\cZ^{KB}(\tau_{n,m/2,0})$\\\hline
\end{tabular}
\end{center}
For Klein Bottle/Annuls/M\"obius strip cases, the moduli
parameter $\tau$ is pure imaginary. $\tau_{n,m,p}$ is
the long string moduli (\ref{e_long_string_moduli}).
$p$ is an integer from $0$ to $n-1$  and $p^*$ is
the half-odd integer from $0$ to $n$.

In this table we used the partition functions
for a single string in each sector. 
We will first prove this table
by employing the explicit operator formalism 
for the simplest target space $\bR^1$.
In this case, they have
the following standard form,
\begin{eqnarray}
 \cZ^T (\tau,\bar\tau) & = & 
\frac{\be{-(\tau-\bar\tau))/24}}{\sqrt{\mbox{Im}\,\tau}}
  \prod_{n=1}^\infty \left((1-\be{n\tau})(1-\be{-n\bar\tau})\right)^{-1}\,\,,
\\
 \cZ^{KB} (\tau) & = & \frac{\be{-\tau/12}}{\sqrt{\mbox{Im}\,\tau}}
  \prod_{n=1}^\infty \left(1-\be{2n\tau}\right)^{-1}\,\,,
\\
 \cZ^A (\tau) & = & \frac{\be{-\tau/24}}{\sqrt{2\mbox{Im}\,\tau}}
  \prod_{n=1}^\infty \left(1-\be{n\tau}\right)^{-1}\,\,,
\\
 \cZ^M (\tau) & = & \frac{\be{-\tau/24}}{\sqrt{2\mbox{Im}\,\tau}}
  \prod_{n=1}^\infty \left(1-(-1)^n\be{n\tau}\right)^{-1}\,\,.
\end{eqnarray}

%We note that there are even some mixing between oriented
%and unorientable surfaces.
%Some of them can be easily explained. For example,
%\begin{enumerate}
% \item In Klein bottle  and M\"obius strip amplitude,
%orientation flip occurs in long string sector when $m$ is even.
%It may be understood as having even number of orientation
%flip in the long string sector. ($m$ is the number of flips
%in this case.)
% \item In annulus and M\"obius strip amplitudes,
%orientation projection naturally occurs when $n$  is even.
%This is a consequence of the fact that the flip change
%can be accomplished by the summation over $p$ when $n$ is even.
%\end{enumerate}

In the first subsection, we give the examination of
this table by using the explicit operator formalism.
In the most cases, we omit the analysis of the momentum integration
since they are the same as our discussion in section 2.

In the second subsection, we present a different
proof based on multiple cover of the  world sheets.
This is similar to our discussion 
in the section \ref{s_general_target} and can be
applied to the arbitrary target space.
Although it may be less rigorous compared to the analysis
by the operator formalism, it is much better to
explain the topological nature of this table.

\subsection{Analysis by Operator formalism}
\subsubsection{Klein Bottle}
As in our calculation in the torus amplitude,
we make the discrete Fourier transformation
(\ref{e_DF}) for the component fields.
The action of $h$ was determined in (\ref{e_hgX}) and
it is possible to use the same mode expansion (\ref{e_ME}).
One novelty is to determine the action of $\inv_n$,
\begin{eqnarray}
 (\inv_n \cdot\tbX)^{a,J} &=& \frac{1}{\sqrt{n}}\sum_{I=0}^{n-1}
\be{-\frac{aI}{n}}\bX^{n-I-1,J}\nn\\
& = & \be{\frac{a}{n}}\tbX^{-a, J}\,\,.
\end{eqnarray}
The action of $g$ in (\ref{e_K}) is evaluated as,
\begin{equation}
 (g\cdot \tbX)^{a,J} = \be{\frac{a p_{J}}{n}}
 \tbX^{-a,J+1}.
\end{equation}
Since the orientation flip interchange the left and the right
movers, $(\Omega\cdot g)$ acts on the oscillators as
\begin{eqnarray}\label{e_Klein_Omega}
 (\Omega\cdot g)\,\alpha_{r-a/n}^{a,J} &= &
  \be{\frac{a p_{J}}{n}}\tilde\alpha_{r-a/n}^{-a,J+1}\,\nn\\
 (\Omega\cdot g)\,\tilde\alpha_{r-a/n}^{-a,J} &= &
  \be{\frac{-ap_{J}}{n}}\alpha_{r-a/n}^{a,J+1}\,.
\end{eqnarray}
We need to find a combination of the left and right movers
which is invariant up to scalar multiplication
under $\Omega\cdot g$. To this end, we define a linear
combination
\begin{equation}
 \beta^{a}_{r-a/n} = \sum_{J=0}^{m-1} C_J \alpha^{a,J}_{r-a/n},
\end{equation}
and define $\tilde{\beta}^a_{r-a/n}\equiv (\Omega\,g)\, 
\beta^{a}_{r-a/n}$.
If one can find appropriate coefficients $C_J$ such that
$\beta$ satisfies
\begin{equation}\label{e_beta_t}
 \Omega\, g\cdot \tilde{\beta}^a_{r-a/n}=\lambda \beta^a_{r-a/n}\,\,,
\end{equation}
$\beta^{a}_{r-a/n}\tilde{\beta}^a_{r-a/n}$ becomes
diagonal under the action of $\Omega\, g$,
\begin{equation}
\Omega\, g \left(\beta^{a}_{r-a/n}\tilde{\beta}^a_{r-a/n}\right)
=\lambda  \left(\beta^{a}_{r-a/n}\tilde{\beta}^a_{r-a/n}\right)\,\,.
\end{equation}
In terms of $C_J$, (\ref{e_beta_t}) becomes
\begin{equation}
 \lambda C_{J+2} = C_J \be{\frac{a}{n}(p_{J}-p_{J+1})}\,\,.
\end{equation}
Since $J$ takes its value in $0,\cdots,m-1$ (modulo $m$),
the solution becomes essentially different depending on 
whether $m$ is even or odd.
\vskip 2mm
\noindent{\bf [1] odd $m$}:

In this case, the recursion relation gives
\begin{eqnarray}
 C_{2m}&=&\frac{1}{\lambda} \be{\frac{a}{n}(p_{2m-2}-p_{2m-1})}C_{2(m-1)}\nn\\
 & = & \cdots\nn\\
 & = & \frac{1}{\lambda^m}\be{
 \frac{a}{n}\left(\sum_{\ell=0}^{m-1} p_{2m-2\ell} -
	     \sum_{\ell=0}^{m-1} p_{2m-1-2\ell}\right)}\,C_0\nn\\
 & = & \frac{1}{\lambda^m}\, C_0\,\,.
\end{eqnarray}
$C_{2m}=C_0$ implies $\lambda=\be{\frac{b}{m}}$ with $b=0,1,\cdots m-1$.

Along the same line of calculation (\ref{e_comb}), we obtain the
oscillator contribution to the partition function of Klein bottle,
\begin{eqnarray}
 \mbox{Tr}_{\cH_h}(\Omega\, g\, \be{\tau (L_0+\bar L_0)})& = & 
  \prod_{s=1}^{\infty} \prod_{a=0}^{n-1}\prod_{b=0}^{m-1}
  \frac{1}{1-\be{\frac{b}{m}+2\tau\left(s-\frac{a}{n}\right)}}\nn\\
 & = &   \prod_{s=1}^{\infty} \prod_{a=0}^{n-1}
  \frac{1}{1-\be{2m\tau\left(s-\frac{a}{n}\right)}}\nn\\
 & = &   \prod_{s=1}^{\infty}
  \frac{1}{1-\be{\frac{2m\tau}{n}}}\,\,.
\end{eqnarray}

The calculation of the zero mode contribution is the same
as the torus.  It proved the table.

\vskip 2mm
\noindent{\bf [2] even $m$}:

In this case, the recursion relation is split into two 
sequences $\left\{C_\ell\right\}_{\ell_{even}}$
and  $\left\{C_\ell\right\}_{\ell_{odd}}$.
The first one can be solved as,
\begin{equation}
C_{m} =\frac{1}{\lambda^{m/2}}\be{
 \frac{ap}{n}}\,C_0\,\,.
\end{equation}
with
\begin{equation}
 p=\sum_{\ell=0}^{m/2-1} p_{m-2\ell} -
	     \sum_{\ell=0}^{m/2-1} p_{m-1-2\ell}\,\,.
\end{equation}
It gives quantized eigenvalues for $\lambda$,
\begin{equation}
 \lambda_{even}^{a,b} = \be{\frac{2b}{m}+\frac{2ap}{nm}}\,\,,
 \quad b=0,\cdots,\frac{m}{2}-1\,\,.
\end{equation}
Similarly, the recursion relation for
$\left\{C_\ell\right\}_{\ell_{odd}}$, gives
\begin{equation}
 \lambda_{odd}^{a,b} = \be{\frac{2b}{m}-\frac{2ap}{nm}}\,\,,
 \quad b=0,\cdots,\frac{m}{2}-1\,\,.
\end{equation}

The calculation of the partition function is now
completely parallel to (\ref{e_comb}),
\begin{eqnarray}
\mbox{Tr}_{\cH_h}(\Omega\, g\, \be{\tau (L_0+\bar L_0)})
 &=&  \prod_{s=1}^{\infty} \prod_{a=0}^{n-1}\prod_{b=0}^{m/2-1}
  \left(1-\be{\frac{2b}{m} -\frac{2ap}{nm}
 +2\tau\left(s-\frac{a}{n}\right)}\right)^{-1}\nn\\
&&\cdot    \left(1-\be{\frac{2b}{m} +\frac{2ap}{nm}
 +2\tau\left(s-\frac{a}{n}\right)}\right)^{-1}\nn\\
&=& \prod_{s=1}^{\infty}  \left(1-\be{\frac{m\tau-p}{n}s}\right)^{-1}
\left(1-\be{\frac{m\tau+p}{n}s}\right)^{-1}
\end{eqnarray}
Together with the zero mode contribution,
this is exactly the same as (\ref{e_Znmp})
except for that the short string moduli $\tau$ is pure imaginary.
Interestingly the moduli for the long string $\tau_{n,m,p}$ 
has the real part as $p/n$.  

%%%%%%%%%%%%%%%%%%%%%%%%%%%%%%%%%%%%%%%%%%%%%%%%%%%%%%%%%%%%%%%%%%
\subsubsection{Annulus}
%%%%%%%%%%%%%%%%%%%%%%%%%%%%%%%%%%%%%%%%%%%%%%%%%%%%%%%%%%%%%%%%%%

\paragraph{$I_A$}
From the data (\ref{e_I_A}), the mode expansion 
that satisfies the boundary conditions (\ref{e_twist_bc})
at the two boundaries is given by,
\begin{eqnarray}\label{e_I_mode_epn}
 \tbX^{a,J}_L & = & i \sum_{s\in\bZ}\frac{1}{s-a/n}\alpha^{a,J}_{s-a/n}
  \be{\sigma_1(-s+a/n)}\nn\\
 \tbX^{a,J}_R & = & i \be{\frac{aq_J}{n}}\sum_{s\in\bZ}
  \frac{1}{s+a/n}\alpha^{-a,J}_{s+a/n}
  \be{\sigma_1(s+a/n)}\,\,.
\end{eqnarray}
The action of $g$ on the other hand is given by,
\begin{equation}
 \left(g\cdot \tbX_L\right)^{a,J} =
  \be{\frac{ap_J}{n}}\tbX^{a,J+1}_L
  = i \be{\frac{ap_J}{n}}\sum_{s\in\bZ}\frac{1}{s-a/n}\alpha^{a,J+1}_{s-a/n}
  \be{\sigma_1(-s+a/n)}\,\,.
\end{equation}
Comparing these two equations, one gets the action of $g$ on
the oscillators,
\begin{equation}\label{e_annul_g}
 (g\cdot \alpha_{s-a/n})^{a,J} = \be{\frac{ap_J}{n}}
 \alpha^{a,J+1}_{s-a/n}\,\,.
\end{equation}
Since this is the same expression that we met in torus amplitude,
one obtain the partition function immediately.

Unlike the torus case, $p=\sum_{J}p_J$ cannot take arbitrary integer.
By summing over $J$ in (\ref{e_I_A_local}), one obtains
$2p\equiv 0$ modulo $n$.  When $n$ is odd, $p=0$ is the
only solution. Putting $p=0$,
one obtains the standard annulus contribution
of one long open string,
\begin{equation}\label{e_part_annul}
 \mbox{Tr}_{\cH_{f_1,f_2}}
  \left(
   g\be{\tau\,L_0}
  \right)
  =
  \prod_{r=0}^\infty \frac{1}{1-\be{\frac{m\tau}{n}r}}\,\,.
\end{equation}

On the other hand, when $n$ is even, we have two solutions,
$p=0,\frac{n}{2}$.  The first one is the annulus. The second
solution gives,
\begin{equation}\label{e_part_mobius}
 \prod_{r=1}^\infty \frac{1}{1+\be{\frac{m\tau}{n}r+\frac{r}{2}}}
 = \prod_{r=1}^\infty \frac{1}{1+(-1)^r\be{\frac{m\tau}{n}r}}.
\end{equation}
This is the standard M\"obius strip amplitude for a long string.

\paragraph{$II_A$ and $\widetilde{II}_A$}

In these cases we write $m=2k$ since $m$ must be even.
The mode expansion and $g$ action is almost the same as 
$I_A$ case,
\begin{eqnarray}\label{e_II_mod_epn}
 \tbX^{a,J}_L & = & i \sum_{s\in\bZ}\frac{1}{s-a/n}\alpha^{a,J}_{s-a/n}
  \be{\sigma_1(-s+a/n)}\nn\\
 \tbX^{a,J}_R & = & i \be{\frac{aq_J}{n}}\sum_{s\in\bZ}
\frac{1}{s+a/n}\alpha^{-a,J+k}_{s+a/n}
  \be{i\sigma_1(s+a/n)}\,\,.
\end{eqnarray}
The action of $g$ 
on the other hand is the same as (\ref{e_annul_g}).
The difference between $II_A$ and $\widetilde{II}_A$ is
the definition of $J$. In $II_A$ (resp. $\widetilde{II}_A$)
case it is defined modulo $2k=m$ (resp. $k$).

In $II_A$ case, the constraint (\ref{e_II_A_local}) gives
$\sum_{J=0}^{2k-1} p_J=0$. The partition function becomes,
\begin{equation}
  \prod_{r=1}^\infty \frac{1}{1-\be{\frac{2k\tau}{n}r}}\,\,,
\end{equation}
which can be identified with Klein bottle amplitude.

In $\widetilde{II}_A$ case, $p=\sum_{J=0}^{k-1}p_J$
can take any value in $0,1,\cdots,n-1$.  The partition function
therefore becomes,
\begin{equation}
  \prod_{r=1}^\infty \frac{1}{(1-\be{\frac{k\tau+p}{n}r})
 ( 1-\be{\frac{k\tau-p}{n}r})}\,\,,
\end{equation}
which gives the torus partition function.

\subsubsection{M\"obius strip}
\paragraph{$I_M$}
The mode expansion is given by (\ref{e_I_mode_epn}).
The action of $g$ together with the orientation flip
is given by,
\begin{eqnarray}
 (\Omega\, g\cdot \tbX)^{a,J}_L(\sigma_1) & = & 
  (g\tbX_R)^{a,J}(\pi-\sigma_1 )\nn\\
 & = & \be{\frac{ap_J}{n}}\tbX^{-a,J+1}_R(\pi-\sigma_1)\nn\\
 & = & \be{\frac{ap_J}{n}}\be{-\frac{aq_{J+1}}{n}}\tbX^{a,J+1}_L(\sigma_1-\pi).
\end{eqnarray}
In terms of the oscillators, it can be evaluated as,
\begin{equation}
 (g\,\Omega\cdot\alpha_{s-a/n})^{a,J}=
  \be{\frac{a}{2n}(2p_J-2q_{J+1}-1)}(-1)^s\alpha_{s-a/n}^{a,J+1}.
\end{equation}
If we denote the eigenvalues of $g\, \Omega$ action as $\mu$,
the same line of argument as in section 2 gives,
$\mu^m=(-1)^{sm}\be{\frac{a}{2n}R}$ with
\begin{equation}
 P=\sum_{J=0}^{m-1} (2p_J-2q_{J+1}-1).
\end{equation}
The constraint (\ref{e_I_M_local}) implies that
$P\equiv 0$ mod $n$.  However, we need to evaluate it
in mod $2n$ to get the accurate phase factor.
The answer is,
\begin{equation}
 P\equiv \left\{
\begin{array}{cl}
 n\qquad& \mbox{$n$,$m$\ odd}\\
 0\qquad& \mbox{$n$ \ odd, $m$\ even}\\
 0, n\qquad& \mbox{$n$,$m$\ even}
\end{array}
\right.
\end{equation}
When $n$ is even and $m$ odd, we do not have any solution to
mod n relation \newline$2\sum_J p_J -2\sum_J q_J -m\equiv 0$.
Although we need some care for $(-1)^s$ factor, 
the rest of the calculation is almost the same. When $P=0$
it gives the annulus partition function (\ref{e_part_annul})
and when $P=n$, it gives M\"obius amplitude,
(\ref{e_part_mobius}).

\paragraph{$II_M$ and $\widetilde{II}_M$}
Let us denote $m=2k$.
The mode expansion is same as $II_A$ cases (\ref{e_II_mod_epn}).
In the similar line of calculation as in $I_M$, one obtains
the action of $\Omega\,g$ on the oscillators,
\begin{equation}\label{e_IIM_g_action}
 \left(\Omega\,g\cdot
 \alpha_{s-a/n}
\right)^{a,J} =
\be{\frac{a}{2n}(2p_{J}-2q_{J+1}-1)}(-1)^s
\alpha^{a,J+k+1}_{s-a/n}.
\end{equation}
The difference between case $II_M$ and $\widetilde{II}_M$
is that
index $J$ in this equation is defined mod $2k$ ($II_M$) or $k$ 
($\widetilde{II}_M$). If we write 
$\alpha^{a,k+J}=\bar\alpha^{a,J}$, the above relation resembles
the Klein bottle case (\ref{e_Klein_Omega}). This is exactly the 
case for $\widetilde{II}_M$. If we use the result in that section,
one gets
\begin{enumerate}
 \item Klein bottle amplitude when $k$ is odd with partition function:
\begin{equation}\label{e_IIM_Klein}
 \prod_{r=1}^\infty \frac{1}{1-\be{\frac{2k\tau}{n}}}\,\,,
\end{equation}
 \item Torus amplitude when $k$ is even,
\begin{equation}
 \prod_{r=1}^\infty \frac{1}{(1-\be{\frac{k\tau+p}{n}r})
(1-\be{\frac{k\tau-p}{n}r})},
\end{equation}
with $p=\sum_{\ell=0}^{k/2-1}(p_{2\ell}+p_{k+2\ell+1})-
\sum_{J=0}^{k-1}q_J-\frac{k}{2}$.
\end{enumerate}

In $II_M$ case, we need to use
``twisted identification'' $\alpha^{a,k}\equiv\tilde\alpha^{a,0}$.
Because of this twist, the correspondence with torus/Klein is
toggled between $k$ being odd/even.

\begin{enumerate}
 \item When $k$ is even, we get the Klein bottle amplitude
(\ref{e_IIM_Klein}).
 \item When $k$ is odd, we get the torus amplitude,
\begin{equation}\label{e_half_twist}
 \prod_{r=1}^\infty \frac{1}{(1-\be{\frac{2k\tau+p}{2n}r})
(1-\be{\frac{2k\tau-p}{2n}r})},
\end{equation}
with $p\equiv2\sum_{\ell=0}^{k-1}p_{2\ell}-2\sum_{J=0}^{k-1}q_J-k$
mod $2n$. We have to note that this quantity is always odd
as long as we evaluate it in mod $2n$.
\end{enumerate}

%%%%%%%%%%%%%%%%%%%%%%%%%%%%%%%%%%%%%%%%%%%%%%%%%%%%%%%%
% derivation by multiple cover 
%%%%%%%%%%%%%%%%%%%%%%%%%%%%%%%%%%%%%%%%%%%%%%%%%%%%%%%%
\subsection{Derivation by multiple cover}
In this subsection, we present the alternative
derivation of the table at the beginning of this section
based on the multiple cover of the short string world sheet.
We first recall that the four sectors are graphically
represented in Figure 6.
%%%%%%%%
 \begin{figure}[ht]
  \centerline{\epsfxsize=12cm \epsfbox{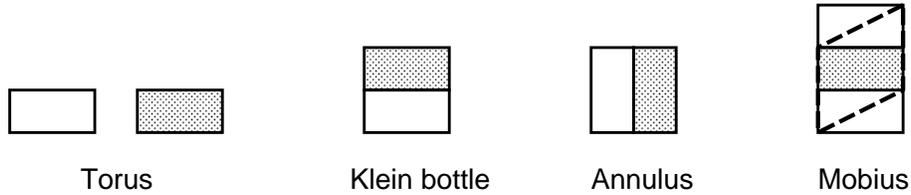}}
  \vskip 3mm
  \caption{World sheets of short strings}
 \end{figure}
%%%%%%%%
Here the white (shaded) box represents the world sheet for the
left (resp. right) mover. In the torus world sheets, the left
and the right movers are independent and are detached from
each other.  In the Klein bottle, two boxes are piled vertically
because of the orientation projection in the trace.
In the annulus, they are piled horizontally
because of the reflections at the boundaries. 
In the M\"obius strip, they are piled horizontally and vertical
at the same time.  The fundamental parallelogram of Figure 4 
is drawn by dashed line.

In the torus case discussed in section \ref{s_general_target},
we first introduce infinite plane with lattice points.
We then assign the name of each short string to each rectangle
by the rule determined from $h$ and $g$.
In the open case, we need to put the left and right movers
in the same plane. In the Klein bottle amplitude,
there are toggles between left/right movers in the vertical
direction (horizontal stripes). In the annulus, we have the
toggling in the horizontal direction (vertical stripes).
In the M\"obius case, we have the toggling in both direction
(checker board). We then put the names of the short
string on each rectangle by using rule from $g$, $h$, and $f$
and  determine the ``fundamental region''.
In the following, we will not try to exhaust all combinations
appearing in the table. Rather we will illustrate the
simple situation which will illuminate the topology change
between the world sheets of the short and the long strings.

\paragraph{Klein bottle}
In this case, there
is toggling between torus/Klein bottle amplitude
when viewed as the long string diagram. We explain
it by taking the simplest situations $n=1$ and $m=2,3$.
%%%%%%%%
 \begin{figure}[ht]
  \centerline{\epsfxsize=8cm \epsfbox{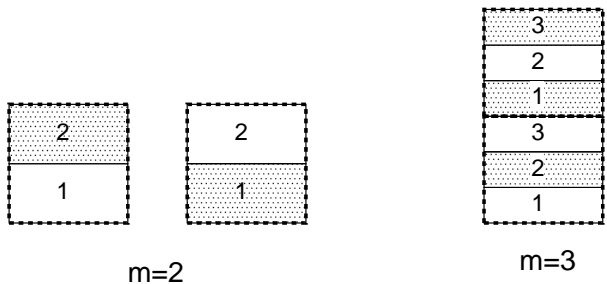}}
  \vskip 3mm
  \caption{Klein bottles for short strings}
 \end{figure}
%%%%%%%%
In these cases, the left (resp. right) mover of $I+1$'th
short string world sheet should be piled over 
the right (resp. left) mover of $I$'th.
When $m=2$, we have independent two vertical loops, 
$L1\rightarrow R2 \rightarrow L1$, 
$R1\rightarrow L2 \rightarrow R1$ where $LI$ ($RI$) means
the world sheet of the left (resp. right) mover
of $I$'th short string. As we illustrate it
in Figure 7 left, we have two independent rectangles of size 2.
From the viewpoint of the long string, they should be
identified as the left and right moving sectors of the
long string. It is easy to observe that 
the similar phenomena occurs whenever $m$ is even.

If $m=3$, the six boxes should be attached with each other
to form one big rectangle. It should be identified as 
the Klein bottle world sheet of the long string.
It is easily generalized that the fat Klein bottle
type world sheet appear whenever $m$ is odd.

Toggling between Annulus/M\"obius strip long string amplitudes
in the $I_M$ sector can be similarly understood.

\paragraph{Annulus}
We first explain the appearance of the long open and closed
strings by taking the simplest example which consists of
two short strings in annulus diagram.  
We have two choices for $f_1$ and $f_2$,
(A) $f_1=(1)(2)$, $f_2=(12)$  (B) $f_1=f_2=(12)$.
%%%%%%%%
 \begin{figure}[ht]
  \centerline{\epsfxsize=14cm \epsfbox{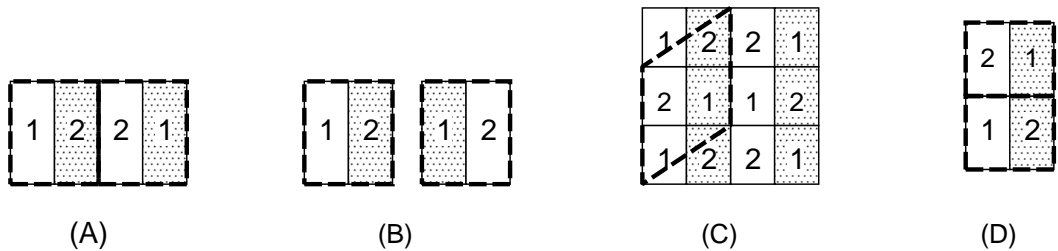}}
  \vskip 3mm
  \caption{Annulus of short strings}
 \end{figure}
%%%%%%%%
In the first case, the horizontal attachments
are defined as in Figure 8 (A) and we have only one chiral
world sheet. This is the situation which describes annulus
diagram for the long string.  In the second case, we have two 
independent groups which are not attached each other
Figure 8 (B). This is again the torus world sheet for
the long string. We therefore meet the world sheet of the long
closed string.

This is not the end of the story.
Even in such a simple situation, we have a degree of freedom
of the twist in the vertical direction. Since we are considering
the annulus diagram, the world sheet of the left (right) mover
should be piled vertically over that of the left (right) mover.
Since we have two boxes for each, we have two choices for $g$,
$g=1$ and $g=(12)$.  In the first case, the diagrams 
is  Figure 8(A,B) themselves.  On the other hand, if we take 
$g=(12)$, we get two new world sheets (C,D).
In diagram (C), as we draw a dashed line, the obtained diagram
should be interpreted as the M\"obius strip for the long string.
For the diagram (D), two independent rectangles in (B) are
piled vertically and gives the Klein bottle world sheet
for the long string.  In the table, (A) and (C) are classified
as $I_A$ with $n=2$, $m=1$. (B) and (D) are classified
as $\widetilde{II}_A$ and $II_A$ respectively with $n=1$ and $m=2$.

In this way, we get all the four diagrams of the long string
amplitude from the annulus of the short string.

\paragraph{M\"obius strip}
In this case, we have the toggling of left and right movers
in both horizontal and vertical directions.
As in the annulus case, we explain the essence of the
correspondence by using two short strip configurations.
In figure 9, we illustrate three possible
configurations.  

%%%%%%%%
 \begin{figure}[ht]
  \centerline{\epsfxsize=14cm \epsfbox{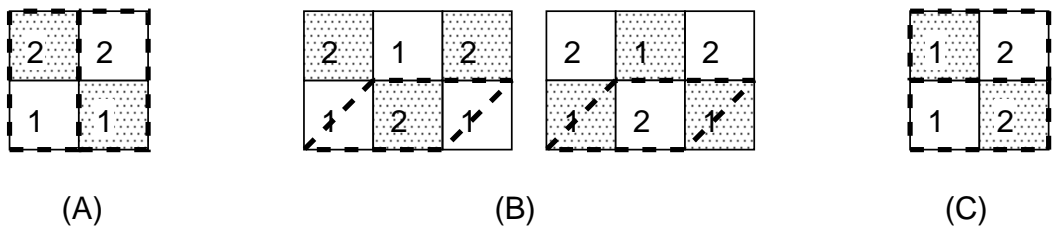}}
  \vskip 3mm
  \caption{M\"obius strip of short strings}
 \end{figure}
%%%%%%%%

The first one (A) corresponds to
$I_M$ case ($n=1, m=2$) with $f_1=(1)(2)$, $g=(12)$.
Because $m$ is even, we get the world sheet of annulus diagram.
The second one corresponds to $II_M$ ($n=1,m=2$) with 
$f_1=g=(12)$. As illustrated in the figure (B), we have two
independent parallelogram region with a twist by one block.
It corresponds to the torus amplitude with $\tau_{1,1,1/2}$.
The appearance of the half an odd integer is the characteristic
feature in $II_M$ case.  
The third one corresponds to $\widetilde{II}_M$ case
with $f_1=(12)$, $g=(1)(2)$. As depicted in the figure (C),
it describes the world sheet of the Klein bottle

%%%%%%%%%%%%%%%%%%%%%%%%%%%%%%%%%%%%%%%%%%%%%%%%%%%%%%%%%%%%%%%%%%
%\newpage
%%%%%%%%%%%%%%%%%%%%%%%%%%%%%%%%%%%%%%%%%%%%%%%%%%%%%%%%%%%%%%%%%%
\section{Boundary states}
%%%%%%%%%%%%%%%%%%%%%%%%%%%%%%%%%%%%%%%%%%%%%%%%%%%%%%%%%%%%%%%%%%
In this section, we show that there are basically 
two types of the boundary/cross-cap states
in the symmetric product orbifold.

The first one is the conventional boundary/cross-cap
states of the long string. One non-trivial
point is that the boundary state for the short
string $\Bket{g}{f}$ sometimes describes the
cross-cap state of the long string. This is one of
the origin of the change of the world sheet
topology in the long string.

The second one describes the joint of two short strings.
It helps to organize arbitrary long string world sheet
from that of the short strings. This is clearly needed
if our orbifold CFT has the character of the second
quantized string theory. We will call such state as
the joint state.

In usual description of the string theory,
the boundary state describes the dynamics of the D-brane.
In our approach, it is naturally unified into the interaction
of the string fields.

\subsection{Boundary states of short strings}
Let us first derive the boundary states for the irreducible
sets. Here we use the terminology ``irreducible'' to mean
that it can not written as the direct product
of the boundary states for the subset fields.

Let us start from a generic irreducible combination which satisfies
$[g,f]=0$,
\begin{equation}\label{e_bits_bc}
 g=diag(\overbrace{\cyclb_{\tn},\cdots,\cyclb_{\tn}}^\tm)\,\,,\qquad
 f=diag(\cyclb_\tn^{\tp_0},\cdots,\cyclb_\tn^{\tp_{\tm-1}})\cdot\cycl_\tm\,\,.
\end{equation}
The condition $f^2=1$ firstly imposes the condition $\tm=1,2$. 
When $\tm=1$ (i.e. $f=\cyclb_\tn^{\tp_0}$), 
 $f^2=1$ implies $2\tp_0\equiv 0$ mod $\tn$.
 If $\tn$ is odd, the only solution is $\tm\equiv 0$
 mod $\tn$. If $\tn$ is even, we have two solutions,
 $\tm\equiv 0,\tn/2$.
When $\tm=2$, $f^2=1$ is satisfied  if $\tp_0+\tp_1\equiv 0$.
We are left with only three types of the boundary states,
\begin{enumerate}
 \item \underline{\em Boundary state of long string}: \ \ 
$(g,f)=(\cyclb_\tn,1)$.  If we use the mode expansion
of the closed string (\ref{e_ME}) (while exchanging $n$ by $\tn$),
we get the explicit expression for the boundary state,
\begin{equation}
 \Bket{\cyclb_n}{1} = \exp\left(
-\sum_{r=1}^\infty \sum_{a=0}^{\tn-1}
\frac{1}{r-a/\tn}\alpha^{(a)}_{-r+a/\tn}
\tilde\alpha^{(\tn-a)}_{-r+a/\tn}
\right)|0\rangle_\tn.
\end{equation}
We introduce the long string oscillator of length $\tn$ as
\begin{equation}
 \cA_{\tn p-a}\equiv \sqrt{\tn}
 \alpha^{(a)}_{p-a/\tn}\,\,,\qquad
 \ctA_{\tn p-a}\equiv \sqrt{\tn}
 \tilde\alpha^{(\tn-a)}_{p-a/\tn}.
\end{equation}
They satisfy a standard commutation relation
$\left[\cA_n,\cA_m\right]=n\delta_{n+m,0}$.
The commutation relation with Hamiltonian
is modified to $\left[L_0,\cA_r\right]=-\frac{r}{\tn}\cA_r$.
In terms of this variable, the boundary state is rewritten as
\begin{equation}
\exp\left(
-\sum_{r=1}^\infty 
\frac{1}{r}\cA_{-r}\ctA_{-r}
\right)|0\rangle_\tn\equiv |B\rangle_\tn\,\, .
\end{equation}
As for the zero mode, since we are considering the
Neumann type boundary condition, we need to impose
(writing $P$ for the momentum for $\cA$)
\begin{equation}\label{e_zero_mode1}
 P|0\rangle_\tn=0\,\,.
\end{equation}
This is nothing but the standard expression of the boundary state.
We will denote the vacuum state for the $J$'s long oscillator satisfying
(\ref{e_zero_mode1})  as $|0_{(J)}\rangle_\tn$ in the following.

\item \underline{\em Cross-cap state for long string}:
\ \ $(g,f)=(\cyclb_{\tn},\cyclb_{\tn}^{\tn/2})$\newline
This state exist only when $\tn$ is even.
(\ref{e_Boundary_state}) is satisfied by
\begin{eqnarray}\label{e_crosscap_state}
\Bket{\cyclb_{\tn}}{\cyclb_{\tn}^{\tn/2}}&\equiv & |C\rangle_\tn \nn\\
 & = &  \exp\left(
-\sum_{r=1}^\infty \sum_{a=0}^{\tn-1}
\frac{(-1)^a}{r-a/\tn}\alpha^{(a)}_{-r+a/\tn}
\tilde\alpha^{(\tn-a)}_{-r+a/\tn}
\right)|0\rangle_\tn \nn\\
& = &\exp\left(
-\sum_{r=1}^\infty 
(-1)^r\frac{1}{r}\cA_{-r}\ctA_{-r}
\right)|0\rangle_\tn\,\, .
\end{eqnarray}
This is again the standard expression for the cross-cap state
of the long string.  This is the origin of the mixture
of annulus/M\"obius amplitude which we have observed in the
previous section.
\item \underline{\em Joint state}:
\begin{equation}
 g=\left(\begin{array}{c c}
    \cyclb_\tn & 0\\ 0 & \cyclb_\tn
 \end{array}\right)\,\,,\qquad
f=\left(\begin{array}{c c}
    0 & \cyclb_\tn^{\tp} \\ \cyclb_\tn^{-\tp} & 0
 \end{array}\right)\,\,.
\end{equation}
This {\em boundary} state actually interconnects two
long strings at the boundary.  The boundary 
condition (\ref{e_Boundary_state}) can be easily solved to give,
\begin{eqnarray}
 \Bket{g}{f}  & \equiv & 
|J \,(12),\tp\rangle_\tn\nn\\
& = &  \exp\left(
\sum_{r=1}^\infty \sum_{a=0}^{\tn-1}
\frac{1}{r-a/\tn}\left(\be{\frac{a\tp}{\tn}}\alpha^{(a,1)}_{-r+a/\tn}
\tilde\alpha^{(\tn-a,2)}_{-r+a/\tn}
\right.\right.\nn\\
&&\left.\left.
+\be{\frac{-a\tp}{\tn}}\alpha^{(a,2)}_{-r+a/\tn}
\tilde\alpha^{(\tn-a,1)}_{-r+a/\tn}
\right)\right)|0\rangle_\tn 
\label{e_Interchange_state}\\
& = &\exp\left(
\sum_{r=1}^\infty 
\frac{1}{r}\left(\be{\frac{r\tp}{\tn}}\cA^{(1)}_{-r}\ctA^{(2)}_{-r}
+\be{\frac{-r\tp}{\tn}}\cA^{(1)}_{-r}\ctA^{(2)}_{-r}\right)
\right)|0\rangle_\tn\nn\,\, ,
\end{eqnarray}
where $\cA^{(I)}$ ($I=1,2$) represents two long string variables.
\end{enumerate}

As for the zero mode, the constraint from Dirichlet type
boundary condition is written in terms of the zero mode as,
\begin{equation}\label{e_zero_mode2}
 (x_0^1-x_0^2)|0\rangle_\tn=(P^1+P^2)|0\rangle_\tn=0\,\,.
\end{equation}
We will denote the vacuum state which satisfy above condition
for $I$th and $J$th long string as $|0_{(IJ)}\rangle$.

\subsection{Cross-cap states of short strings}
%As we have seen, the boundary and the cross-cap states
%for the long string are already prepared in the boundary states
%of the string bits. For the completeness of our discussion,
%let us write down the explicit form of the 
%cross-cap states for the short strings.

In the cross-cap states, $g$ and $f$ satisfies $f^2=g$.
Since $\left[g,f\right]=\left[f^2,f\right]=0$, 
$g$ and $f$ should be written in the form (\ref{e_bits_bc}).
As in the previous subsection, the constraint $f^2=g$
implies $\cyclb_\tm^2=1$. So the only possibility
is $\tm=1,2$.  

When $\tm=1$, $f=\cyclb^\tp$ and $f^2=g$
gives $2\tp\equiv 1$ mod $\tn$.  If $\tn$ is odd, we have
one solution $\tp=(\tn+1)/2$. On the other hand if $\tn$ is even
we have no solution.  
When $\tm=2$, the general solution to $f^2=g$ is
(\ref{e_bits_bc}) with $p_0+p_1\equiv 1$ mod $n$.
We end up with two class of solutions,
\begin{enumerate}
 \item \underline{\em Cross-cap states of the long string}:
$g=\cyclb_\tn$ and $f=\cyclb_\tn^{(\tn+1)/2}$
for odd $\tn$. \newline
In this case one may reshuffle the basis to write $f=\cyclb_\tn$
and $g=\cyclb_\tn^2$.
The mode expansion (\ref{e_ME}) is slightly modified to
\begin{equation}\label{e_ME2}
 \widetilde\bX^{a}=\alpha_0\delta_{a,0}\sigma_0
+i\sum_{r\in\bZ}\left(
\frac{1}{r-2a/\tn}\alpha^{a}_{r-2a/\tn}z^{-r+2a/\tn}+
\frac{1}{r+2a/\tn}\tilde\alpha^{a}_{r+2a/\tn}\bar{z}^{-r-2a/\tn}
\right)
\end{equation}
The boundary condition (\ref{e_crosscap}) can be written 
in terms of oscillators as
\begin{equation}
 (\alpha^{a}_{r-2a/\tn}+(-1)^r\tilde\alpha^{\tn-a}_{-r+2a/\tn})
\Cket{f} =0\,\,,
\end{equation}
where the twist factor $f$ (\ref{e_crosscap}) is exactly 
canceled by the translation of $\sigma_1$.
Now it is straightforward to write down the cross-cap
state,
\begin{equation}
\Ckets{\cyclb_{\tn}^2}{\cyclb_{\tn}}  =   \exp\left(
-\sum_{r=1}^\infty \sum_{a=0}^{\tn-1}
\frac{(-1)^r}{r-2a/\tn}\alpha^{(a)}_{-r+2a/\tn}
\tilde\alpha^{(\tn-a)}_{-r+2a/\tn}
\right)|0\rangle_\tn \nn\\
\end{equation}
It is not difficult from this point to show 
that after redefinition of 
the long string variable, it reduces to (\ref{e_crosscap_state}).

 \item \underline{\em Joint state}:
\begin{equation}
 g=\left(\begin{array}{c c}
    \cyclb_\tn & 0\\ 0 & \cyclb_\tn
 \end{array}\right)\,\,,\qquad
f=\left(\begin{array}{c c}
    0 & \cyclb_\tn^{\tp+1} \\ \cyclb_\tn^{-\tp} & 0
 \end{array}\right)\,\,.
\end{equation}
After repeating the similar calculation, we arrive at
(\ref{e_Interchange_state}) with $\tp$ replaced by
$\tp+1/2$. Clearly this is the origin of half-twist
which appeared in the fat torus amplitude (\ref{e_half_twist}).
\end{enumerate}

In either cases, the condition for the zero-mode are
the same as in the boundary states 
(\ref{e_zero_mode1},\ref{e_zero_mode2}).

%%%%%%%%%%%%%%%%%%%%%%%%%%%%%%%%%%%%%%%%%%%%%%%%%%%%%%%%%%%%%%%%%%
\subsection{Inner product between boundary states}
%%%%%%%%%%%%%%%%%%%%%%%%%%%%%%%%%%%%%%%%%%%%%%%%%%%%%%%%%%%%%%%%%%
In this subsection, we calculate the inner product
between boundary states of the short strings
to see the modular property of the long open strings.
In the following, we will restrict our attraction
to the boundary states since the calculation
of the cross-cap states are completely analogous.

In order to reproduce the open string amplitudes
from the boundary state, we need to loosen the 
the irreducibility of the boundary state.
We therefore start from the general form (\ref{e_bits_bc})
and impose the condition $f^2=1$. 

The general solution
is given by
\begin{equation}\label{e_general_f}
 f=diag(\cyclb^{\tp_0}_\tn,\cdots,\cyclb_\tn^{\tp_{\tm-1}})\cdot
\cycl_{\tm}^q\inv_\tm\,\,.
\end{equation}
The constraint $f^2=1$ further imposes
\begin{equation}\label{e_tp_cons}
 \tp_\ell +\tp_{\tm-1-\ell+q}\equiv 0\quad\mbox{mod} \,\, \tn\,\,.
\end{equation}
The corresponding boundary state $\Bket{g}{f}$
can be decomposed into the product of the (long string)
boundary, cross-cap, and joint states.
When 
\begin{equation}\label{e_loose_end}
\ell\equiv\tm-1-\ell+q\qquad \mbox{ mod}\,\,  \tm\,\,, 
\end{equation} 
it is described by
the boundary and cross-cap states and otherwise it is 
given by the joint state.  Let us first count the
number of the boundary and cross-cap states.

%%%%%%%%
 \begin{figure}[ht]
  \centerline{\epsfxsize=8cm \epsfbox{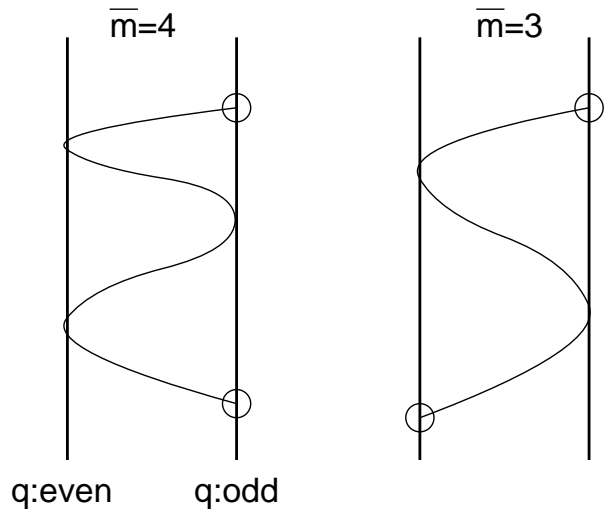}}
  \vskip 3mm
  \caption{Loose ends of long string}
 \end{figure}
%%%%%%%%

We remark first that $\tm$ is the number of long open strings
which have their open end at the boundary.
When $\tm$ is even and $q$ is odd, we have two solutions to
(\ref{e_loose_end}), $\ell = (q-1)/2, (q-1+\tm)/2$.
In this case, the long string has two loose end at that boundary
and the others are connected each other
(Figure 10 left).
When $\tm$ is even and $q$ is even, we have no solution and
every long strings are jointed each other.
On the other hand, when $\tm$ is odd, we always have only 
one loose end (Figure 10 right).

\subsubsection{Annulus/M\"obius strip/Klein bottle amplitudes}
Let us assume that $f_1$ and $f_2$ (reflection factors
at each boundary) to have the general form (\ref{e_general_f})
\begin{equation}
 f_1=diag(\cyclb^{\tp^{(1)}_0}_\tn,\cdots,\cyclb_\tn^{\tp^{(1)}_{\tm-1}})\cdot
\cycl_{\tm}^{q_1}\inv_\tm\,\,,\quad
 f_2=diag(\cyclb^{\tp^{(2)}_0}_\tn,\cdots,\cyclb_\tn^{\tp^{(2)}_{\tm-1}})\cdot
\cycl_{\tm}^{q_2}\inv_\tm\,\,.
\end{equation} 

Annulus, M\"obius strip and Klein bottle amplitudes
can be obtained if there are two loose ends at the boundaries.
Such situations can be obtained if 
(i) $\tm$ is odd, or (ii) $\tm$ is even and $q_1-q_2$  is odd.
To get the irreducible diagram, one may always 
obtain a basis where $q_1=1$ and $q_2=0$ in either cases.

In the explicit evaluation, we use the following formula,
\begin{equation}
 \langle 0 | e^{-f_2^{IJ}\alpha_I\tilde\alpha_J q}
e^{-f_1^{KL}\alpha^\dagger_K\tilde\alpha^\dagger_L}|0\rangle
=\prod_{a}\frac{1}{1-\mu_a q}\,\,,
\end{equation}
where oscillators satisfy the commutation relations
$\left[\alpha_I,\alpha^\dagger_J\right]
=\left[\tilde\alpha_I,\tilde\alpha^\dagger_J\right]
=\delta_{IJ}$ 
and $\mu_a$ are the eigenvalues of $f_2 f_1^t$.

In our case, $f_2\,f_1^t=f_2 \,f_1$ is given by
\begin{equation}
 diag(\cyclb_\tn^{\tp_0^{(2)}+\tp_{\tm-1}^{(1)}}
,\cyclb_\tn^{\tp_1^{(2)}+\tp_{\tm-2}^{(1)}},\cdots)\cdot
\cycl_{\tm}^{-1}.
\end{equation}

As we have been doing in section 2, we first diagonalize
the $\cyclb_\tn$ action by discrete Fourier transformation.
In the subspace 
where $\cyclb_\tn \cdot \bX^{a,J}=\be{\frac{a}{\tn}}\bX^{a,J}$,
the eigenvalues of this matrix should satisfy
\begin{equation}
 \mu^\tm =\be{\frac{a}{\tn}\left(\sum_{\ell=0}^{\tm-1}
(\tp^{(2)}_\ell + \tp^{(1)}_{\tm-1-\ell})
\right)}
\end{equation}
Because of (\ref{e_tp_cons}), almost all the terms
except for $\tp$ at the loose ends cancel each other in
the right hand side. For the annulus/Klein bottle amplitude,
the phase at the loose end cancel each other and $\mu^\tm=1$.
For the M\"obius strip amplitude, we have non-trivial phase
$\mu^\tm=(-1)^a$ since $\sum\tp=\tn/2$.

For the oscillator contribution of the 
annulus and Klein bottle amplitude, we obtain
the inner product as,
\begin{eqnarray}\label{e_A_from_BS}
 \Bbra{g}{f_2}q^{L_0+\bar{L}_0}\Bket{g}{f_1}^{oscillator}& = &
 \prod_{r=1}^\infty \prod_{a=0}^{\tn-1}\prod_{b=0}^{\tm-1}
 \frac{1}{1-\be{b/\tm}q^{2(r-a/\tn)}}\nn\\
 & = &  \prod_{r=1}^\infty 
 \frac{1}{1-q^{2\tm r/\tn}}
\end{eqnarray}
Similarly for the M\"obius strip amplitude ($\tn$ should
be  even in this case),
\begin{eqnarray}\label{e_M_from_BS}
 \Bbra{g}{f_2}q^{L_0+\bar{L}_0}\Bket{g}{f_1}^{oscillator}& = &
 \prod_{r=1}^\infty \prod_{a=0}^{\tn-1}\prod_{b=0}^{\tm-1}
 \frac{1}{1-\be{\frac{b}{\tm}+\frac{a}{2\tm}}q^{2(r-a/\tn)}}\nn\\
 & = &  \prod_{r=1}^\infty 
 \frac{1}{1-(-1)^r q^{2\tm r/\tn}}
\end{eqnarray}

For the zero mode contribution to the 
inner product, we use the momentum
representation of the vacuum states,
\begin{equation}
 \langle P^I | 0_{(I)}\rangle_\tn = \delta(P^I),
\qquad
 \langle P^I, P^J | 0_{(IJ)}\rangle_\tn = \delta(P^I+P^J).
\end{equation}
If $\tm$ is even, the inner product can be written 
in the following form,
\begin{eqnarray}
&& {}_\tn\langle 0_{(1)(23)\cdots(\tm-2,\tm-1)(\tm)}|
q^{\frac{1}{2\tn}(L_0+\bar L_0)}
|0_{(12)\cdots(\tm-1,\tm)}\rangle_\tn\nn\\
&&\qquad =
%\frac{1}{(2\pi)^\tm}
\int d^\tm P\,\,
{}_\tn\langle 0_{(1)(23)\cdots(\tm-2,\tm-1)(\tm)}|\vec{P}\rangle
q^{\frac{1}{2\tn}\sum_I P_I^2}
\langle \vec{P}|0_{(12)\cdots(\tm-1,\tm)}\rangle_\tn\nn\\
&&\qquad =
\int d^\tm P\,\,
\delta(P_1)
\delta(P_1+P_2)\delta(P_2+P_3)\cdots\delta(P_{\tm-1}+P_\tm)
\delta(P_\tm)
q^{\frac{1}{2\tn}\sum_I P_I^2}
\nn\\
&&\qquad = \delta(0) =V.
\label{e_inner_zero_mode}
\end{eqnarray}
Here $V$ is the volume and $|0_{(12)\cdots(\tm-1,\tm)}\rangle_\tn$ is the
short hand notation of $|0_{(12)}\rangle_\tn\otimes\cdots\otimes
|0_{\tm-1,\tm}\rangle_\tn$.
%$V$ must be absorbed in the normalization of the ground state
%and we put the result to be 1.
Calculation for odd $\tm$ is mostly the same and gives
the same answer.

\subsubsection{Torus}
In order to get the torus amplitude, $\tm$ should be
even since the both boundary should have no loose ends.
Similarly both $q_1$ and $q_2$ are even to fulfill this relation.
In order to get the irreducible amplitude, one may put $q_1=2$ 
and $q_2=0$ without losing generality.
In this case $f_2f_1$ is given by
\begin{equation}
 diag(\cyclb_\tn^{\tp_0^{(2)}+\tp_{\tm-1}^{(1)}}
,\cyclb_\tn^{\tp_1^{(2)}+\tp_{\tm-2}^{(1)}},\ldots)\cdot
\cycl_{\tm}^{-2}.
\end{equation}
Since $\tm$ is even, we have two sets of eigenvalue equations,
\begin{equation}
 \mu^{\tm/2} =\be{\pm\frac{a\tp}{\tn}},\quad
\tp\equiv\sum_{\ell=0}^{\tm/2-1}
(\tp^{(2)}_{2\ell} + \tp^{(1)}_{\tm-1-2\ell})\,\,.
\end{equation}
Unlike the situation in the previous subsection, 
$\tp$ can take any integer.

Writing $k=\tm/2$, we obtain the oscillator part of the torus amplitude,
\begin{eqnarray}
&& \Bbra{g}{f_2}q^{L_0+\bar{L}_0}\Bket{g}{f_1}^{oscillator}  \nn\\
&&\qquad = \prod_{r=1}^\infty \prod_{a=0}^{\tn-1}\prod_{b=0}^{k-1}
 \frac{1}{(1-\be{\frac{b}{\tk}+\frac{a\tp}{\tk}}q^{2(r-a/\tn)})
(1-\be{\frac{b}{\tk}-\frac{a\tp}{\tk}}q^{2(r-a/\tn)})}\nn\\
  &&\qquad =  \prod_{r=1}^\infty 
 \frac{1}{(1-\be{\frac{2\tk\tau+\tp}{\tn}})(1-\be{\frac{2\tk\tau-\tp}{\tn}})}
\end{eqnarray}

For the zero mode calculation, the calculation is
parallel to (\ref{e_inner_zero_mode}). The only difference
is the product of the delta functions of $P$. The integral
produces,
\begin{eqnarray}
&&\int d^\tm P\,\,
\delta(P_1+P_2)\delta(P_2+P_3)\cdots\delta(P_{\tm-1}+P_\tm)
\delta(P_{\tm}+P_1)
q^{\frac{1}{2\tn}\sum_I P_I^2}
\nn\\
&& \qquad = \delta(0)\int dP q^{\frac{\tm}{2\tn} P^2}
 =  V (\mbox{Im}\,\tau_{\tn,\tm,0})^{-1/2} \,\,.
\end{eqnarray}
The moduli dependence appearing here is necessary
to have the correct modular property.

%%%%%%%%%%%%%%%%%%%%%%%%%%%%%%%%%%%%%%%%%%%%%%%%%%%%%%%%%%%%%%%%%%
%\newpage
%%%%%%%%%%%%%%%%%%%%%%%%%%%%%%%%%%%%%%%%%%%%%%%%%%%%%%%%%%%%%%%%%%
\section{DLCQ partition function}
%%%%%%%%%%%%%%%%%%%%%%%%%%%%%%%%%%%%%%%%%%%%%%%%%%%%%%%%%%%%%%%%%%

In this section, we prove that the various amplitudes
of the open string are organized to give the simple formula
similar to (\ref{e_exponential_formula}).
Such a structure is needed to identify
our model as the DLCQ  of the open string
field theory as we reviewed in section \ref{DLCQ_review}.

Unlike the closed  string partition function, the requirement of 
the modular invariance does not necessarily determine
the combinations of the twisted open string sectors
(\ref{e_annulus}, \ref{e_Mobius}).
It is quite encouraging that the summation over the 
all possible twists indeed gives the DLCQ type
partition function.

Usually in BCFT, the combination of various sectors
is determined by the tadpole cancellation condition.
In terms of the boundary state and the cross cap state,
it is written as the cancellation of the massless part of
the boundary states,
\begin{equation}\label{e_tadpole}
 |B\rangle\!\rangle_0 + |C\rangle\!\rangle_0=0\,\,.
\end{equation}
In the bosonic string in the flat target space, this condition
determine the gauge group should be $SO(2^{13})$.
In other situations, this gives a crucial constraint on the
model building.
While we try to apply the tadpole condition (\ref{e_tadpole})
naively,
we meet one difficulty. Namely the
generic one loop amplitude is reducible and its irreducible
components have a tachyonic part. Multiplying them
usually produces the higher negative modes
which complicate the constraint (\ref{e_tadpole}). 
It usually becomes a nonlinear relation and the analysis
would be very difficult.
Such a situation will be remedied if we have the DLCQ
type partition function. In this case,
products of the various string amplitudes
can be organized as the exponential of the sum of
the irreducible ones. We can examine the 
tadpole condition for each of the single long string 
amplitudes.

We will summarize our results in the following theorem,

%In this section, we first prove the analog of
%theorem 1 in the open string amplitudes.
%We will then prove the tadpole condition 
%leads to the conventional gauge group $SO(2^{13})$.

\vskip 5mm
\noindent{\bf Theorem 3 :}
\vskip 3mm
{\em
\noindent (i) \underline{Klein bottle}\footnote{
While we are typing this manuscript, we noticed 
the work \cite{KLEIN-BOTTLE} where authors independently derived
the partition function of Klein bottle amplitude.
}:
\newline
The generating function of the partition function,
\begin{equation}
 Z^{KB}_N(\tau) =\frac{1}{N!} 
\sum_{g,h\in S_N}\mbox{Tr}_{\cH_h}
(g\Omega^{closed}\be{\tau (L_0+\bar{L}_0)}),
\end{equation}
with a constraint $hg=gh^{-1}$, is written as
\begin{eqnarray}\label{e_DLCQ_partition_KB}
&& \ln(\sum_{N=0}^\infty \zeta^N Z_N^{KB}(\tau))\\
&&\quad = 
\sum_{n,m=1}^\infty \left(\frac{\zeta^{n(2m-1)}}{2m-1} 
\cZ^{KB}(\tau_{n,2m-1})
+\frac{\zeta^{2nm}}{2nm}\sum_{p=0}^{n-1}\cZ^{Torus}
(\tau_{n,2m,p},\bar\tau_{n,2m,p})\right)
.\nn
\end{eqnarray}
\vskip 3mm
\noindent (ii) \underline{Annulus} : \newline
The generating function of the partition function,
\begin{equation}
 Z_N^{Annulus}=\frac{1}{N!}\sum_{h, g,f_1,f_2\in S_N}
\mbox{Tr}_{\cH_{f_1,f_2}}
\left(g\,\be{\tau L_0}\right)\,\,,
\end{equation}
with the constraint,
$
 \left[g,f_1\right]= \left[g,f_2\right]=0
$,
$f_1^2=f_2^2=1$, $h=f_2^{-1}f_1$
is written in the following form,
\begin{eqnarray}\label{e_DLCQ_partition_A}
 &&\ln(\sum_{N=0}^\infty \zeta^N Z_N^{Annulus}(\tau))\nn\\
 &&\quad =\sum_{n,m=1}^\infty \frac{\zeta^{nm}}{m}
  \cZ^{Annulus}(\tau_{n,m,0})+
\sum_{n,m=1}^\infty \frac{\zeta^{2nm}}{m}
  \cZ^{Mobius}(\tau_{2n,m,0})\\
 &&\quad+\sum_{n,m=1}^\infty \frac{\zeta^{2nm}}{2nm}
 \sum_{p=0}^{n-1}\cZ^{Torus}(\tau_{n,m,p},\bar\tau_{n,m,p})
 +\sum_{n,m=1}^\infty \frac{\zeta^{2nm}}{2m} \cZ^{KB}(\tau_{n,m,0}).\nn
\end{eqnarray}
\vskip 3mm
\noindent (iii) \underline{M\"obius strip} :\newline
The generating function of the partition function,
\begin{equation}
 Z_N^{Mobius}=\frac{1}{N!}\sum_{h, g,f\in S_N}
\mbox{Tr}_{\cH_{f,gfg^{-1}}}
\left(g\Omega^{open}\,\be{\tau L_0}\right)\,\,,
\end{equation}
with the constraint,
$ h=gfg^{-1}f$, $f^2=1$, $ \left[g^2,f\right]= 0$,
is given as follows,
\begin{eqnarray}
 &&\ln(\sum_{N=0}^\infty \zeta^N Z_N^{Mobius}(\tau))\nn\\
 &&\quad =\sum_{n,m=1}^\infty \frac{\zeta^{2nm}}{2m}
  \cZ^{Annulus}(\tau_{n,2m,0})+
\sum_{n,m=1}^\infty \frac{\zeta^{(2n-1)m}}{m}
  \cZ^{Mobius}(\tau_{2n-1,m,0})
\label{e_DLCQ_partition_M}\\
 &&\quad+\sum_{n,m=1}^\infty \frac{\zeta^{2nm}}{2nm}
 \sum_{p=0}^{n-1}\cZ^{Torus}(\tau_{n,m,p+n/2},\bar\tau_{n,2m,p+n/2})
 +\sum_{n,m=1}^\infty \frac{\zeta^{2nm}}{2m} \cZ^{KB}(\tau_{n,m,0}).\nn
\end{eqnarray}
}
\vskip 5mm

%%%%%%%%%%%%%%%%%%%%%%%%%%%%%%%%%%%%%%%%%%%
%%  Proof
%%%%%%%%%%%%%%%%%%%%%%%%%%%%%%%%%%%%%%%%%%
\noindent{\bf Proof:}\newline
{\em Klein bottle:}
The strategy is completely parallel to our 
discussion in  section \ref{s_GF}.
We first choose an element $h$ which
represents each conjugacy class. For each $h$, we need
to count the number of the
elements $g$ which satisfies this constraint. 
By comparing (\ref{e_hg}) and (\ref{e_K}), 
we have the same degree of freedom for $g$
including the introduction of parameter $p$. The only deference
is that the Klein bottle amplitude at the end does not depend
on $p$. Therefore the summation $\frac{1}{n}\sum_p$ 
does not exist. We can conclude that it has the same type
of the  generating functional (\ref{e_DLCQ_partition_KB}).
\vskip 3mm
%%%%%%%%%%%%%%%%%%%%%%%%%%%%%%%%%%%%%%%%
\noindent{\em Annulus and M\"obius strip:}\newline
We have three class of solutions 
(\ref{e_I_A},\ref{e_II_A},\ref{e_IIt_A}).  
For given $N$, the annulus partition function is
the sum of the products of all possible combinations of
three irreducible solutions. If we combine them
into the generating function, the contributions
from each diagram are factorized and we can 
count them independently.  Although we have extra
summation over $f_i$, the number of the solutions 
are almost the same if we look at 
(\ref{e_I_A},\ref{e_II_A},\ref{e_IIt_A}) carefully.

For type $I_A$ solutions, there are one constraint
on $p$, $2\sum_\ell p_\ell\equiv 0$. For each value of
$\sum_\ell p_\ell$, we have $n$ solutions for $q$ which satisfies
(\ref{e_I_A_local}). Since the final expression (\ref{e_part_annul})
does not depend on $q$, we have the same number of 
degree of freedom as in the torus case.
This type of counting holds exactly the
same fashion for type $II_A$ solutions.

For type $\widetilde{II}_A$, we need to be more careful
to count the combinations. However, the final expression
turns out to be the same as in $II_A$ case.

By combining all types of the solutions, we arrive at 
(\ref{e_DLCQ_partition_A}).

The calculation of the combinatorics 
is exactly the same for the M\"obius strip case.
\qed
\vskip 5mm

%%%%%%%%%%%%%%%%%%%%%%%%%%%%%%%%%%%%%%%%%%%%%%%
%\newpage
%%%%%%%%%%%%%%%%%%%%%%%%%%%%%%%%%
%  Tadpole Condition
%%%%%%%%%%%%%%%%%%%%%%%%%%%%%%%%%
\section{Tadpole Condition}
In the previous section, we have seen that
four sectors (Torus, Klein bottle, Annulus, M\"obius strip)
of the {\em long} string are actually contained in the 
annulus diagram of the {\em short} open string.
This leads us to suspect that we may accomplish
the tadpole condition (\ref{e_tadpole}) by combining the
boundary states of the short strings.
One subtle issue is that
there is no cross-cap state when $\tn$ is odd
(it is described as the cross-cap state of the short string).
In order to achieve the tadpole condition in terms of boundary states
alone, we need to restrict the length of the long strings  ($\tn$)
to be  even.  This condition may be imposed by setting
normalization factor for those boundary states is zero.

In the following, we will restrict our discussion
to the tadpole cancellation between various amplitudes
for the {\em single} long string.  It is 
not necessarily equivalent to the tadpole condition
in conventional BCFT \cite{HARVEY-MINAHAN}--\cite{ISHIBASHI-ONOGI}.
Usually the tadpole condition is used to derive consistent
compactification of the target space.  In our case, however,
the symmetric product is used to describe the string field
theory.  We believe that the our treatment is the 
appropriate one in this physically different context.

Before we start the discussion,
we should comment on the dimension of the target space.
In the following we will discuss on the standard bosonic
string. In order to recover the Lorentz covariance, we need put the 
transverse dimension of the target space to be 24.
The change of the various amplitude is straightforward.

As was seen in section 6.3, the boundary and cross-cap states 
of long strings are constructed by the products of 
irreducible boundary, cross-cap and joint states.
The normalizations of these states should be also factorized
into those of  each irreducible boundary states. 
We will denote them as $\kappa_{B},\kappa_{C}$ and $\kappa_{J}$.
%So we should find the tadpole condition for the irreducible 
%normalization factors.
The structure of the boundary and cross-cap states 
of the long strings are slightly different 
if $\tm$ is even or $\tm$ is odd.

First, we will discuss odd $\tm$ case (Figure 10 right).
There is one loose end on each boundary.
Long string boundary states are written as 
\begin{equation}
  \Bket{g}{f^{B}} = \kappa_{J}^{(\tm -1)/2}\kappa_{B}
  \sum_{\left\{\tp\right\}}
(\prod_{i=1}^{(\tm-1)/2}|J(2i-1,2i),\tp_{i}\rangle_\tn )
    \otimes  |B\rangle_\tn\,\, ,
\end{equation}
where
$g=diag(\overbrace{\cyclb_{\tn},\cdots,\cyclb_{\tn}}^\tm)\,\,$.
$f^B=diag(\cyclb^{\tp^{(B)}_0}_\tn,\cdots,
\cyclb_\tn^{\tp^{(B)}_{\tm-1}})\cdot
\cycl_{\tm}^{q_B}\inv_\tm\,$.
It contains a loose end (boundary) on $\tm$'th long string.
Long string cross-cap states are written as 
\begin{equation}
  \Bket{g}{f^{C}} = \kappa_{J}^{(\tm -1)/2}\kappa_{C}
  \sum_{\left\{\tp\right\}}
 (\prod_{i=1}^{(\tm-1)/2}|J(2i-1,2i),\tp_{i}\rangle_\tn  )
    \otimes  |C\rangle_\tn\,\, ,
\end{equation}
where 
$f^C=diag(\cyclb^{\tp^{(C)}_0}_\tn,\cdots,\cyclb_\tn^{\tp^{(C)}_{\tm-1}})\cdot
\cycl_{\tm}^{q_C}\inv_\tm\,\,$. It contains a cross-cap 
on $\tm$'th long string.

Tadpole cancellation condition \ref{e_tadpole} 
is factorized into the following conditions,
\begin{eqnarray}
&& \kappa_{B}|B\rangle_{\tn 0}\,\, 
   + \kappa_{C}|C\rangle_{\tn 0}\,\,=0\,\,.
\label{e_cond_BC}
\\
&& \sum_{\tp_i} |J(2i-1,2i),\tp_i\rangle_{\tn 0}=0\,\,.
\label{e_cond_I}
\end{eqnarray}
The condition for the joint states (\ref{e_cond_I}) is 
satisfied automatically since,
\begin{eqnarray}
&& \sum_{\tp_i} |J(2i-1,2i),\tp_i\rangle_{\tn 0}\nn\\
&& \qquad = \left(\sum_p \be{p/\tn}\right)(\cA^{(2i-1)}_{-r}\ctA^{(2i)}_{-r}
+\cA^{(2i)}_{-r}\ctA^{(2i-1)}_{-r})|0_{2i-1,2i}\rangle
=0\,\,.
\end{eqnarray}
There are no constraints for  $\kappa_{J}$.
%becomes 
%\begin{eqnarray}
% & &   \Bket{g}{f^{B}}_0  +  \Bket{g}{f^{C}}_0  \nn \\
% &&\qquad =  \kappa_{I}^{(\tm-1)/2} (\prod_{i=1}^{(\tm-1)/2} 
%      |J \,e_i,\tp_i\rangle_{\tn 0} )
%  \otimes (\kappa_{B}|B\rangle_{\tn 0}\,\, 
%   + \kappa_{C}|C\rangle_{\tn 0}\,\, ),
% \nn \\
% &&\qquad =    0 .  
%\end{eqnarray}
For the boundary and the cross-cap states, tadpole condition 
is satisfied when
\begin{equation} \label{irred_tad}
 \kappa_{B}-\kappa_{C}=0 \,\,.
\end{equation}

Let us move to even $\tm$ case (Figure 10 left).
There are two loose ends on one boundary and no loose ends on the other.
The long string boundary states (written as $\Bket{g}{f^{J}}$) 
which does not have any loose ends, 
don't make any contributions for the tadpole cancellation conditions
since they are factorized into the joint states.

We introduce the long string boundary state 
for Annulus $\Bket{g}{f^{A}}$,  M\"obius strip $\Bket{g}{f^{MS}}$,
and Klein bottle $\Bket{g}{f^{KB}}$ as follows,
\begin{eqnarray}
   \Bket{g}{f^{A}}& =& \kappa_{J}^{(\tm -2)/2}\kappa_{B}^2
    (\prod_{i=1}^{\tm /2-1}\sum_{\tp^{i}}
|J(2i,2i+1),\tp_{i}\rangle_\tn )
    \otimes  |B(1)\rangle_\tn\,\, \otimes  |B(\tm)\rangle_\tn\,\, ,\nn\\
   \Bket{g}{f^{MS}}& =& \kappa_{J}^{(\tm -2)/2}\kappa_{B}\kappa_{C}
(\prod_{i=1}^{\tm /2-1}\sum_{\tp_{i}}
|J(2i,2i+1),\tp_{i}\rangle_\tn )\nn\\
&&\qquad    \otimes  (|B(1)\rangle_\tn\,\, \otimes |C(\tm)\rangle_\tn\,\, 
             +|C(1)\rangle_\tn\,\, \otimes |B(\tm)\rangle_\tn\,\,),\\
\Bket{g}{f^{KB}} &=& \kappa_{J}^{(\tm -2)/2}\kappa_{C}^2
(\prod_{i=1}^{\tm /2-1}\sum_{\tp_{i}}
|J(2i,2i+1),\tp_{i}\rangle_\tn )
    \otimes  |C(1)\rangle_\tn\,\, \otimes  |C(\tm)\rangle_\tn\,\, .\nn
\end{eqnarray}
The  name of these states comes from the inner product
formulae,
\begin{eqnarray}
 Z^{A} &\sim&
 \Bbra{g}{f^{J}}q^{L_0+\bar{L_0}}\Bket{g}{f^{A}}\,\,,
 \nn\\
 Z^{M} &\sim&
 \Bbra{g}{f^{J}}q^{L_0+\bar{L_0}}\Bket{g}{f^{MS}}\,\,,\nn\\
 Z^{KB}& \sim&
 \Bbra{g}{f^{J}}q^{L_0+\bar{L_0}}\Bket{g}{f^{KB}}.
\end{eqnarray}
Tadpole cancellation condition (\ref{e_tadpole}) becomes 
\begin{eqnarray}
 &&  \Bket{g}{f^{A}}_0+\Bket{g}{f^{MS}}_0+\Bket{g}{f^{KB}}_0 \nn \\
 && \qquad =   \kappa_{J}^{(\tm -2)/2}
( (\prod_{i=1}^{\tm /2-1}\sum_{\tp_i}|J(2i,2i+1) 
\tp_{i} 
\rangle_{\tn})
\\
 & & \left.\quad\qquad \otimes (\kappa_{B} |B(1)\rangle_{\tn}\,\,
   + \kappa_{C} |C(1)\rangle_{\tn})
\otimes (\kappa_{B} |B(\tm)\rangle_{\tn}\,\,
   + \kappa_{C} |C(\tm)\rangle_{\tn})\right)_0=0\,\,.\nn
\end{eqnarray}
Since it is factorized, it does not produce any new constraints
on $\kappa$'s.

%%%%%%%%%%%%%%%%%%%%%%%%%%%%%%%%%%%%%%%%%%%%%
%    Modular property
%%%%%%%%%%%%%%%%%%%%%%%%%%%%%%%%%%%%%%%%%%%%%

To determine Chan-Paton factor, we calculate
the modular properties of various open string amplitudes.
In section 6 we determine the inner products between various
boundary states.  Together with the normalization factors,
annulus/M\"obius strip/Klein bottle amplitudes
with length $\tm$ are given as (\ref{e_A_from_BS}, \ref{e_M_from_BS}),
\begin{eqnarray}
 \mbox{Annulus :} & \quad & (\tn\kappa_J)^{\tm-1}(\kappa_B)^2
 \be{\frac{\tm}{\tn\tau}}\prod_{r=1}^\infty \left(1-\be{-\frac{\tm}{\tn\tau}r}
\right)^{-24}\,\,,\nn\\
 \mbox{M\"obius :} & \quad & 2(\tn\kappa_J)^{\tm-1}(\kappa_B)^2
 \be{\frac{\tm}{\tn\tau}}\prod_{r=1}^\infty 
\left(1-(-1)^r\be{-\frac{\tm}{\tn\tau}r}
\right)^{-24}\,\,\nn\\
 \mbox{KB :} & \quad & (\tn\kappa_J)^{\tm-1}(\kappa_B)^2
 \be{\frac{\tm}{\tn\tau}}\prod_{r=1}^\infty 
\left(1-\be{-\frac{\tm}{\tn\tau}r}
\right)^{-24}\,\,.
\end{eqnarray}
To achieve the tadpole condition (\ref{e_cond_BC}), we need to restrict $\tn$
to be even. We will write $\tn=2\tk$ in the following.

After the modular transformation, these amplitudes are rewritten as,
\begin{eqnarray}
 \mbox{Annulus :} & \quad & (\tn\kappa_J)^{\tm-1}(\kappa_B)^2
\left(\frac{2\tk\tau}{i\tm}\right)^{-12}
\be{-\frac{2\tk\tau}{\tm}}\prod_{r=1}^\infty \left(1-\be{\frac{2\tk\tau}{\tm}r}
\right)^{-24}\,\,,\nn\\
 \mbox{M\"obius :} & \quad & 2(\tn\kappa_J)^{\tm-1}(\kappa_B)^2
\left(\frac{\tk\tau}{i\tm}\right)^{-12}
\be{-\frac{\tk\tau}{2\tm}}\prod_{r=1}^\infty \left(1-
(-1)^r\be{\frac{\tk\tau}{2\tm}r}
\right)^{-24}\,\,,\nn\\
 \mbox{KB :} & \quad & (\tn\kappa_J)^{\tm-1}(\kappa_B)^2
\left(\frac{2\tk\tau}{i\tm}\right)^{-12}
\be{-\frac{2\tk\tau}{\tm}}\prod_{r=1}^\infty \left(1-\be{\frac{2\tk\tau}{\tm}r}
\right)^{-24}\,\,.\label{e_expr_from_BS}
\end{eqnarray}
These expression should be compared to the partition functions
obtained in section 5.  
\begin{eqnarray}\label{e_expr_from_open}
 \mbox{Annulus :} & \quad & \frac{N^2}{4}
\left(\frac{2m\tau}{in}\right)^{-12}
\be{-\frac{m\tau}{n}}\prod_{r=1}^\infty \left(1-\be{\frac{m\tau}{n}r}
\right)^{-24}\,\,,\nn\\
 \mbox{M\"obius :} & \quad & N\eta
\left(\frac{m\tau}{ik}\right)^{-12}
\be{-\frac{m\tau}{2k}}\prod_{r=1}^\infty \left(1-
(-1)^r\be{\frac{m\tau}{2k}r}
\right)^{-24}\,\,,\nn\\
 \mbox{KB :} & \quad & 
\left(\frac{\tau k}{in}\right)^{-12}
\be{-\frac{2k\tau}{n}}\prod_{r=1}^\infty \left(1-\be{\frac{2k\tau}{n}r}
\right)^{-24}\,\,.
\end{eqnarray}
Here $N$ is the Chan-Paton factor for the long open strings
and we write $m=2k$ in M\"obius and KB amplitudes
since they appear only when $m$ is even.
By comparing expressions, we first need impose
$\kappa_J=\tn^{-1}$ since there are no length dependent
factors in (\ref{e_expr_from_open}).
By comparing the oscillators, we need to impose,
\begin{itemize}
 \item Annulus: $m=2\tk$, $n=\tm$
 \item M\"obius: $m=\tk$, $k=\tm$
 \item Klein bottle: $k=\tk$, $n=\tm$
\end{itemize}
It shows that we need to project to even $m$ sector for
Annulus in (\ref{e_DLCQ_partition_A}) because 
the restriction that $\tn$  is even.
For other sectors, it reproduces every terms in 
(\ref{e_DLCQ_partition_A}).
By comparing the normalization factor, we get
\begin{eqnarray}
 \frac{1}{4}N^2 2^{-12} & = & (\kappa_B)^{2}\,\,\,,\nn\\
 N\eta & = & 2(\kappa_B)^2\,\,,\nn\\
 1 & = & 2^{-12} (\kappa_B)^2\,\,.
\end{eqnarray}
It produces, $\kappa_B=2^{6}$, $\eta=1$ and $N=2^{13}$.
We thus have the standard gauge group $SO(2^{13})$ 
for the bosonic string.

%\newpage
%%%%%%%%%%%%%%%%%%%%%%%%%%%%%%%%%%%%%%%%%%%%%%%%%%%%%%%%%%%%%%%%%%%%
\section{Discussion}
%%%%%%%%%%%%%%%%%%%%%%%%%%%%%%%%%%%%%%%%%%%%%%%%%%%%%%%%%%%%%%%%%%%%
As we mentioned in the introduction,
one of the main goal of the current project is to construct
the second quantized open string theory which has the powerful
handling of D-brane.  For this purpose, 
we studied the detailed combinatorial aspects of the open 
matrix string theory. 
We hope that our argument is convincing enough that
the theory have quite reasonable structure as
the second quantized open string theory.

One distinct merit of current approach to conventional
string field theory is the description of D-brane.
From the boundary conformal field theoretical
viewpoint, the classification
of the possible boundary states should be
interpretable as the possible geometric configuration of 
D-branes (for example see \cite{RECKNAGEL-SCHOMERUS}--\cite{BDLR}).
In our formalism, it is very straightforward to include various
D-brane configurations as the description of the
loose ends of the long open strings.  They can be deformed
by introducing the marginal transformations of the short strings.

On the other hand, in the string field theory, 
we need the information of the D-branes in the very definition
of the string fields. Introduction of several D-branes
may force us to introduce new string fields 
and the string vertex operators for each of D-branes.
Although this approach is useful in the calculation of the tachyon 
condensation \cite{SEN}--\cite{BSZ},
the string field theory may not give
an economic description of the multiple D-brane background.

This approach is also economical in the description
of the string interaction vertices.  As already mentioned in
\cite{JOHNSON}, there is only one open string vertex operator
$\Phi_{KL}$ which interchanges  $K$'th and $L$'th open
strings at the boundary. In terms of the boundary states,
this operator mixes the boundary/cross-cap states
and the joint states,
\begin{eqnarray}
 |J(KL)\rangle &\leftrightarrow& |B(K)\rangle \otimes |B(L)\rangle\nn\\
 |J(KM)\rangle\otimes |J(LN)\rangle&\leftrightarrow&
 |J(LM)\rangle\otimes |J(KN)\rangle.\label{e_interaction}
\end{eqnarray}
%%%%%%%%
 \begin{figure}[ht]
  \centerline{\epsfxsize=10cm \epsfbox{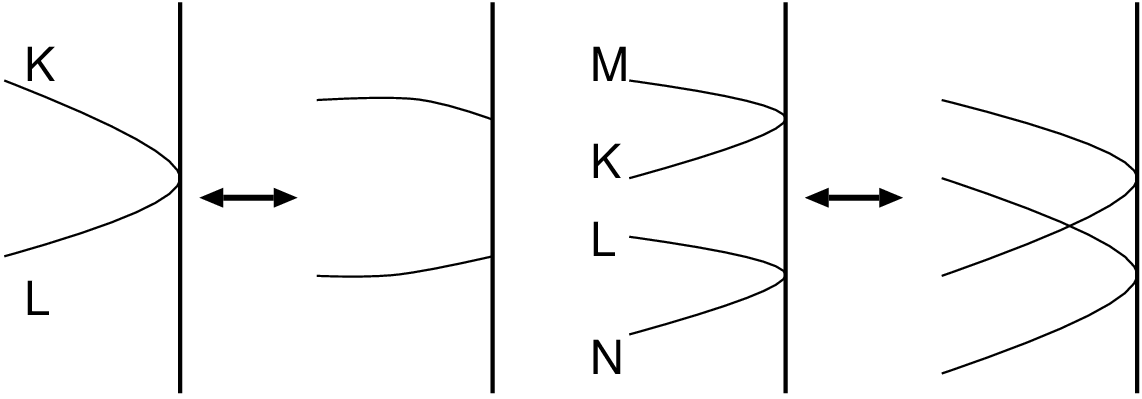}}
  \vskip 3mm
  \caption{Open string interaction}
 \end{figure}
%%%%%%%%
In the open/closed string field theory, we need to introduce
seven types of the string interaction vertices
\cite{KUGO-ASAKAWA-TAKAHASHI}. This is because the
the global topology of interaction vertex becomes
rather involved and we need the vertex operator for
each of them.  In the matrix string approach, 
all of these vertices can be described in terms of one
vertex $\Phi_{KL}$ \cite{JOHNSON}. This is again a
great benefit of the current formalism. 

In case of the string field theory, the gauge invariance of
the string fields requires the gauge group should
be $SO(2^{13})$ \cite{KUGO-ASAKAWA-TAKAHASHI}.
Although we have not attempted the consistency 
of the vertex operator, it should reproduce the similar
condition.  This is one of the most important issues
which should be clarified in the future.

In our discussion in section 5, we emphasized that
there are no big difference between the boundary/cross-cap
states and the joint states. They are 
three equally possible boundary states from the viewpoint
of the short strings. This aspect is clearer
in the action of the interaction vertex (\ref{e_interaction})
since it mixes three boundary states.
While the boundary states are the representation 
of the D-brane, the joint states represents
a kind of string interaction.  In this way, we have seen
an interesting mixture of the string dynamics and D-brane.
The consistency of the interaction $\Phi_{KL}$ will
impose the possible deformation of the joint states
from the knowledge of the D-branes and vice versa.

In this paper, we do not make an explicit attempt
to incorporate the supersymmetry. 
A straightforward
generalization of the closed string \cite{DVV}
was already made in \cite{JOHNSON}.
These aspects are using the same type of the combinatorics 
as the bosonic situation does not produce
extra non-triviality except for the Chan-Paton factor.
One of the difficult point which we would like to indicate 
is that these vertex operators
should intertwine Ramond-Ramond boundary states which describe
the D-brane charge and the joint states.

Finally we have to mention that current development
of the matrix string theory \cite{WYNTER}--\cite{BRAX}
where the nontrivial world sheet topology is interpreted
as the instanton sectors of 2D Yang-Mills theory.
It is quite interesting to investigate if there are similar
description of the open string world sheet as the topologically
non-trivial sectors in the Yang-Mills theory.

\vskip 5mm
\noindent{\bf Acknowledgement:}
{\em
The authors would like to thank
N. Ishibashi for sending us his thesis \cite{ISHIBASHI2}
which was the essential resource for us to understand the BCFT
on orbifolds. 
One of the authors (Y.M.) is obliged to T. Kawai for the information
on DLCQ type partition function and to  T. Kawano for
explaining the work \cite{KUGO-ASAKAWA-TAKAHASHI}.

Y.M is  supported in part by Grant-in-Aid ($\sharp$09640352)
and in part by  Grant-in-Aid for Scientific Research in a Priority
Area: ``Supersymmetry and Unified Theory of Elementary Particle''
($\sharp$707) from the Ministry of Education, Science, Sports and Culture.
}

%\newpage 

\end{document}